\documentclass[journal]{IEEEtran}

\ifCLASSINFOpdf
\usepackage[pdftex]{graphicx}
\DeclareGraphicsExtensions{.pdf,.jpeg,.png}
\else
\usepackage[dvips]{graphicx}
\fi
\usepackage{subfigure}
\usepackage{epstopdf}
\usepackage[cmex10]{amsmath}
\usepackage{amssymb}
\usepackage{wasysym}
\usepackage{relsize}
\usepackage{algorithm}
\usepackage{algorithmic}
\usepackage{array}
\usepackage{cite}
\usepackage{color}
\usepackage{url}
\usepackage{bm}
\usepackage{enumerate}
\usepackage{epsfig,latexsym}
\usepackage{cite}

\usepackage{hyperref}

\hypersetup{
	colorlinks=true,
	linkcolor=blue,
	citecolor=blue
}

\begin{document}
%
\title{Near-Field Multiuser Communications based on Sparse Arrays}

\author{Kangjian~Chen,~\IEEEmembership{Student~Member,~IEEE}, Chenhao~Qi,~\IEEEmembership{Senior~Member,~IEEE}, \\    Geoffrey Ye Li,~\IEEEmembership{Fellow,~IEEE} and Octavia A. Dobre,~\IEEEmembership{Fellow,~IEEE}
	\thanks{This work was supported in part by the National Natural Science Foundation of China under Grants U22B2007 and 62071116, in part  by the  National Key Research and Development Program of China under Grant 2021YFB2900404, in part by the Natural Sciences and Engineering Research Council of Canada (NSERC) through its Discovery program, in part by the SEU Innovation Capability Enhancement Plan for Doctoral Students under Grant CXJH\_SEU 24088. (\textit{Corresponding author: Chenhao~Qi})}
	\thanks{Kangjian~Chen and Chenhao~Qi are with the School of Information Science and Engineering, Southeast University, Nanjing 210096, China (e-mail: \{kjchen, qch\}@seu.edu.cn).}
	\thanks{Geoffrey Ye Li is with the Department of Electrical and Electronic Engineering, Imperial College London, SW7 2AZ London, U.K. (e-mail: geoffrey.li@imperial.ac.uk).}
	\thanks{Octavia A. Dobre is with the Faculty of Engineering and Applied Science, Memorial University, St. John’s, NL A1C 5S7, Canada (e-mail: odobre@mun.ca).}
}

\markboth{Accepted by IEEE Journal of Selected Topics in Signal Processing}
{}
\maketitle

\begin{abstract}
This paper considers near-field multiuser communications based on sparse arrays (SAs). First, for the uniform SAs (USAs), we analyze the beam gains of channel steering vectors, which shows that increasing the antenna spacings can effectively improve the spatial resolution of the antenna arrays to enhance the sum rate of multiuser communications.  Then, we investigate nonuniform  SAs (NSAs) to mitigate the high multiuser interference from the grating lobes of the USAs. To maximize the sum rate of near-field multiuser communications, we optimize the antenna positions of the NSAs, where a successive convex approximation-based antenna position optimization algorithm is proposed. Moreover, we find that the channels of both the USAs and the NSAs show uniform sparsity in the defined surrogate distance-angle (SD-A) domain. Based on the channel sparsity, an on-grid SD-A-domain orthogonal matching pursuit (SDA-OMP) algorithm is developed to estimate multiuser channels. To  further improve the resolution of the SDA-OMP, we also design an off-grid SD-A-domain iterative super-resolution channel estimation algorithm. Simulation results demonstrate the superior performance of the proposed methods.

\end{abstract}


\section{Introduction}
Next-generation wireless communications are expected  to improve user experience and support compelling applications. These anticipated  developments entail comprehensively heightened demands on future communications, such as enhanced transmission rates and improved network connectivity. To  meet these demands, various novel technologies have been developed~\cite{OPCS23NBY}. Among them, near-field communications have attracted  widespread attention  for the  potential to improve the spatial resolution, communication capacity and transmission security~\cite{CM23ZHY,Tcom22CMH,OJCS23LYW,ACSSC21BE,Tcom23LY,TWC23CKJ,TWC23SX}.

According to the array aperture and the propagation distance, the radiation field of the electromagnetic  (EM) waves can be divided into the far field, the radiative near field, and the reactive near field~\cite{CM23ZHY,Tcom22CMH,OJCS23LYW,ACSSC21BE,Tcom23LY}. In the far field, where the propagation distance is sufficiently large, the EM  waves exhibit a planar wavefront. As the propagation distance decreases and the radiative near field is reached, the  EM waves  show a spherical wavefront and the phase differences among antennas are nonlinear functions of the antenna indices. In the reactive near field, where the propagation distance is even closer than that of the radiative near field, the amplitudes of EM waves vary across the array in addition to the nonlinear phases. Since the radiative near field usually has a much larger coverage than the reactive near field~\cite{Tcom22CMH,OJCS23LYW,ACSSC21BE},  we mainly focus on the radiative near field in this work.


Due to the distinctions in propagation characteristics, the near- and far-field channels are described  by the spherical- and planar-wave models, respectively.  By exploiting the unique propagation characteristics in the near-field, various improvement can be achieved compared to far-field communications~\cite{WCL23JR,JSAC23WZD,TVT23CA,TVT24ZZ}. For example, in near-field data transmission, the single-user  communications can benefit from the increased channel degrees of freedom (DoF) to improve system capacity~\cite{WCL23JR}  while the  multiuser  communications can exploit the increased spatial resolution to serve users at the same angle but different distances~\cite{JSAC23WZD}. Since the EM waves exhibit a spherical wavefront,  the near-field steering vector depends on both the angle and distance, enabling the near-field sensing and localization~\cite{TVT23CA}. Moreover, for the near-field physical-layer security, the BS can use the near-field beamforming to simultaneously provide high beamforming gains for legitimate users while low beamforming gains for eavesdroppers even if they share the same angle~\cite{TVT24ZZ}. The improvement indicates that near-field communications can provide  superior  communication capacity and broader applications compared to the far-field ones.





In fact, the implementation of the near-field communications relies on the near-field effects, which are characterized by the spherical wavefront  of the EM waves and   caused by large array apertures. To form large-aperture arrays, most of the existing works utilize the extremely large-scale multiple-input multiple-output (XL-MIMO), which employs much  more antennas than the conventional massive MIMO.  However, this approach entails exorbitant hardware costs. To reduce the hardware costs, in \cite{Tcom22YLF} and \cite{CC2023LXR}, the widely-spaced multi-subarray (WSMS) architecture is developed, where the entire array is divided into several subarrays and the spacings between adjacent subarrays are widened to increase the array aperture. Although this approach outperforms the conventional massive MIMO, its adaptability is constrained because only the spacings between  subarrays can be adjusted and the design flexibility is limited. Different from the existing XL-MIMO and WSMS, we expand the array aperture by increasing the spacings between adjacent antennas. In fact, this idea coincides with the concept of sparse arrays (SAs)~\cite{Access21AI,AWPL22YSJ,TVT20WXH,TAP14GV}.  Compared to the existing XL-MIMO, the SAs enable us to exploit the near-field effects for performance improvement with much lower hardware costs. Compared to the existing WSMS, the SAs increase the spacings between adjacent antennas instead of between subarrays and thus have more design flexibility.

Many studies have been conducted on SAs~\cite{Access21AI,AWPL22YSJ,TVT20WXH,TAP14GV}. For the far field, through adjusting spacings between antennas, the SAs can improve the system performance and reduce the hardware costs~\cite{Access21AI,AWPL22YSJ}. For example, better DoA estimation performance can be achieved~\cite{Access21AI} and fewer antennas are needed to synthesize the same beams by the SAs~\cite{AWPL22YSJ} than the conventional half-wavelength arrays. Besides, some researchers have also explored  the potential of SAs in the near field. For example, the near-field localization based on sparse cross arrays  and the sparse array optimization for  near-field imaging  are investigated in~\cite{TVT20WXH} and \cite{TAP14GV}, respectively. Despite these efforts, to the best knowledge of the authors, so far there has been no work reporting the near-field communications based on SAs.


In this paper, we consider near-field multiuser communications based on SAs. By expanding the array aperture via increasing the antenna spacings, the spatial resolution of the antenna arrays is improved and the hardware costs are reduced.  The main contributions of this paper are summarized as follows, where the second point is included in the conference paper~\cite{SPAWC24CKJ}.

\begin{itemize}
 	\item We investigate near-field multiuser communications based on uniform SAs (USAs) and analyze the beam gains of channel steering vectors. Based on the analysis, we highlight three unique properties of the USA channels, which shows that enlarging the antenna spacings can effectively enhance the spatial resolution of the antenna arrays and improve the sum rate of multiuser communications.
 	
 	\item Then, we investigate nonuniform SAs (NSAs) to mitigate the high multiuser interference from the grating lobes of USAs. To maximize the sum rate of near-field multiuser communications, we optimize the antenna positions of the NSAs in the antenna panel. Since the antenna position optimization problem is nonconvex, a successive convex approximation-based antenna position  optimization (SCA-APO) algorithm is proposed.  
 	
 	\item We explore the channel sparsity of both the USAs and the NSAs. We find that  the channels of the USAs show uniform and periodic sparsity in the defined surrogate distance-angle (SD-A) domain while the channels of the NSAs show uniform and aperiodic sparsity in the SD-A domain.  Based on the channel sparsity, channel sparse representation matrices are designed for the USAs and the NSAs, respectively. Then, an SD-A-domain orthogonal matching pursuit (SDA-OMP) algorithm is proposed to estimate the multiuser channels.  To further improve the resolution of the SDA-OMP, an SD-A-domain iterative super-resolution channel estimation (SDA-ISRCE) algorithm is further proposed.
\end{itemize}

\begin{figure}[!t]
	\centering
	\includegraphics[width=75mm]{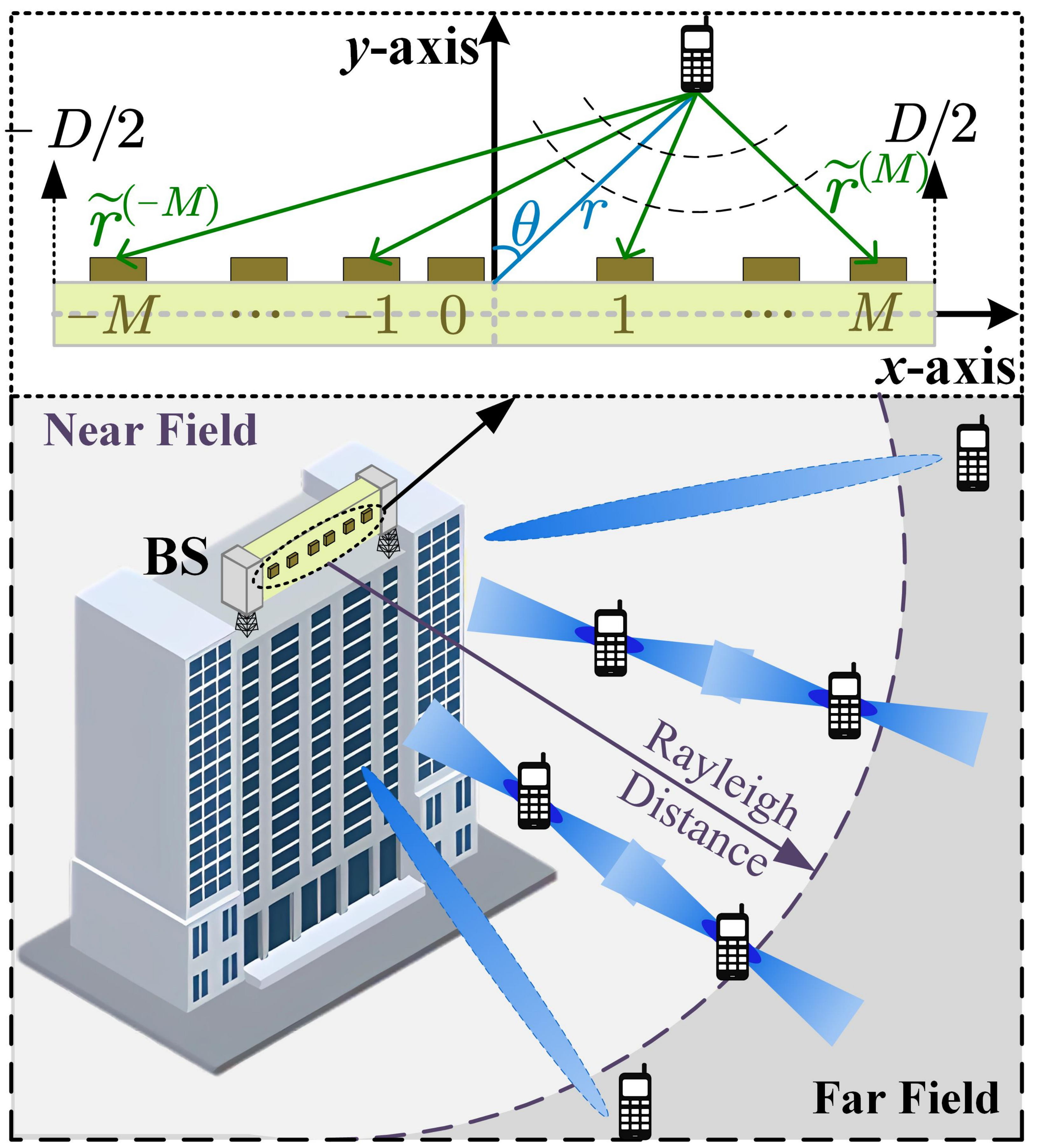}
	\caption{Illustration of the system model.}
	\label{MultipathChannelModel}
\end{figure}

The rest of this paper is organized as follows: The model of near-field multiuser communications based on sparse arrays is introduced in Section~\ref{SystemModel}. The analysis of the USAs is presented in Section~\ref{MCUSLA}. The antenna position optimization of the NSAs is proposed in Section~\ref{MUNSLA}. The channel estimation and beamforming are discussed in Section~\ref{CEP}. The proposed methods are evaluated in Section~\ref{SimulationResults}. The paper is concluded in Section~\ref{Conclusion}.

The notations are defined as follows: Symbols for matrices (upper case) and vectors (lower case) are in boldface. $(\cdot)^{\rm H}$ denotes the conjugate transpose (Hermitian). $[\boldsymbol{a}]_{n}$ represents the $n$th entry, $\left[ \boldsymbol{A} \right]_{:,n}$ denotes the $n$th column, and $\left[ \boldsymbol{A} \right]_{m,n}$ refers to the entry at the $m$th row and $n$th column of matrix $\boldsymbol{A}$. Additionally, $j$ is the square root of $-1$, $|\cdot|$ is the absolute value of a scalar, $|\cdot|_{\rm F}$ is the Frobenius norm of a matrix, $\mathbb{E}$ denotes the expectation operation, $\mathbb{C}$ is the set of complex numbers, $\mathbb{Z}$ is the set of integers, and $\mathcal{C}\mathcal{N}$ represents the complex Gaussian distribution. Furthermore, $f'(\cdot)$ and $f''(\cdot)$ represent the first-order and the second-order derivatives of $f(\cdot)$, respectively.


\section{System Model}\label{SystemModel}
As shown in Fig.~\ref{MultipathChannelModel}, we consider the uplink transmission between $K$ users and a BS. The BS  employs an antenna panel with a length of $D$ to accommodate an $N$-element sparse linear array. To simplify the expressions, we assume $N$ is an odd number so that $M\triangleq(N-1)/2$ is an integer. However, the proposed methods can be extended to antenna arrays with an even number of elements.  To fully exploit the spatial DoF of near-field communications, the fully digital structure is adopted at the BS, which implies that each antenna is connected to a radio frequency chain. For the SAs, due to the much larger antenna spacing than the conventional half-wavelength-interval uniform linear array (HULA), the number of antennas can be much smaller for a fixed antenna panel. Therefore, the budget of RF chains for SAs with fully digital structure can be affordable. For uplink channel estimation, the $K$ users transmit orthogonal pilots to the BS\footnote{Usually, the near-field communications are expected to support more users than the far-field ones. In this condition, the BS may not be able to allocate orthogonal pilots to all users for channel estimation. In our future work, we will investigate the pilot sharing techniques to address this issue~\cite{TWC21WJ}.}. Therefore, received signals from  the $K$ users can be effectively separated at the BS. In this work, we focus on the processing at the BS and assume that users are equipped with only one antenna for simplicity. However, the proposed methods can be extended to the scenarios with multi-antenna users. Then, the received signals from the $k$th user, for $k=1,2,\cdots,K$, can be expressed as 
\begin{align}\label{SysModel}
	\boldsymbol{y}_k = \boldsymbol{h}_k z_k + \boldsymbol{\eta},
\end{align}
where $\boldsymbol{h}_k\in\mathbb{C}^{N}$ represents the channel between the $k$th user and the BS, and $z_k$ denotes the transmit pilot of the $k$th user. $\boldsymbol{\eta}$ denotes the additive white Gaussian  noise and follows $\boldsymbol{\eta}\sim\mathcal{CN}(0,\sigma^2\boldsymbol{I})$. We establish a Cartesian coordinate system to characterize the channels, where the tangent direction, normal direction, and center of the antenna panel are designated as the x-axis, the y-axis, and the origin, respectively. Naturally, the left and right boundaries of the antenna panel are $-D/2$ and $D/2$, respectively. Denote the coordinate of the $n$th antenna, for $n=-M,\cdots,0,\cdots,M$, as $(x_n,0)$. Typically, antenna positions are restricted to the boundaries of the antenna panel, i.e., $x_n \in [-D/2,D/2]$. To intuitively show the sparsity of the SAs, we define  $p\triangleq\frac{2D}{(N-1)\lambda}$, which denotes  the ratio of the length of the antenna panel to the array aperture of the $N$-element HULA. For simplicity, we refer to ``$p$" as the array sparsity factor.

In the radiative near field, according to the uniform spherical-wave model~\cite{OJCS23LYW}, the channel between the $k$th user and the BS can be expressed as
\begin{align}\label{ChannelModel}
	\boldsymbol{h}_k = \sum_{l=1}^{L_k}\gamma_k^{(l)} \boldsymbol{\alpha}\left(\boldsymbol{x},r_k^{(l)},\theta_k^{(l)}\right),
\end{align}
where $L_k$ and $\boldsymbol{x}\triangleq [x_{-M},\cdots,x_0,\cdots,x_M]^{\rm T}$ denote the number of paths between the $k$th user and the BS, and the stack of the antenna positions, respectively. $\gamma_k^{(l)}$, $r_k^{(l)}$, and $\theta_k^{(l)}$ denote the channel gain,  channel distance,  and channel physical angle-of-departure (AoD) of the $l$th path  between the $k$th user and the BS, respectively. $\gamma_k^{(l)}$ follows the complex Gaussian distribution with a mean of zero and a variance of $\xi_k^{(l)}$. In \eqref{ChannelModel}, we assume equal path loss from all antennas in the radiative near field.  The accuracy of this approximation may degrade for  non-broadside directions, which are usually not the interested directions of the BS due to the severe gain degradation~\cite{mailloux1994phased}. Therefore, the spherical-wave model in \eqref{ChannelModel} would be accurate in the  radiative near field under a reasonable application environment. $\boldsymbol{\alpha}\left(\boldsymbol{x},r_k^{(l)},\theta_k^{(l)}\right)$, which is a function of $\boldsymbol{x}$, $r_k^{(l)}$  and $\theta_k^{(l)}$, denotes the channel steering vector for the  $l$th path between the $k$th user and the BS. We omit the superscript and subscript in $r_k^{(l)}$ and $\theta_k^{(l)}$ for simplicity and express $\boldsymbol{\alpha}\left(\boldsymbol{x},r,\theta\right)$~as 
\begin{align}\label{ChannelSteeringVector}
	\big[\boldsymbol{\alpha}\left(\boldsymbol{x},r,\theta\right)\big]_n = e^{j2\pi\left(\widetilde{r}^{(n)} -r\right)/\lambda},
\end{align}
for $n=-M,\cdots,0,\cdots,M$, where $\lambda$ denotes the carrier wavelength. $\widetilde{r}^{(n)}$ represents the distance between the $n$th antenna of the SA and the user and can be expressed as 
\begin{align}\label{Dist}
	\widetilde{r}^{(n)} = \sqrt{r^2-2rx_n\sin\theta + x_n^2}.
\end{align}

The complex expression in \eqref{Dist} poses great difficulties to the system implementation and performance analysis. To simplify the expression, we approximate $\widetilde{r}^{(n)}$ by 
\begin{align}\label{SecOrdTayExpansion}
	\widetilde{r}^{(n)} \approx r-x_n\sin\theta + \frac{x_n^2(1-\sin^2\theta)}{2r},
\end{align}
according to $\sqrt{1+\epsilon}\approx1+\epsilon/2-\epsilon^2/8$, which is verified to be accurate in the radiative near field, i.e., $r\ge0.62\sqrt{D^3/\lambda}$~\cite{OJCS23LYW}. Considering a communication system with $D=1$~m and $\lambda = 0.01$~m, we have $0.62\sqrt{D^3/\lambda} = 6.2$~m, which is much smaller than the typical coverage of the BS. Therefore, in this work, we mainly focus on the radiative near field with $r\ge0.62\sqrt{D^3/\lambda}$. Define $\Theta\triangleq \sin\theta $ and 
\begin{align}\label{surrogatedistance}
b\triangleq\frac{(1-\Theta^2)}{2r}.
\end{align}
Note that $b$ is a distance-dependent function.  Therefore, we refer to ``$b$" as the ``surrogate distance". In addition, $\Theta$ represents the angle information of channel paths. Therefore, we refer to ``$\Theta$" as the ``angle" for simplicity.  Since the physical AoD usually satisfies $\theta \in [-90^\circ,90^\circ]$, we have $\Theta\in [-1,1]$. When $r$ is very large, $b$  is close to zero. When $\Theta$ equals zero and  $r$ equals the minimum distance of the BS coverage, $b$ achieves its maximum value of $b_{\rm max}$. Therefore, we have $b\in[0,b_{\rm max}]$.  Substituting \eqref{SecOrdTayExpansion} and \eqref{surrogatedistance} into \eqref{ChannelSteeringVector}, the channel steering vector can be simplified~as 
\begin{align}\label{SimplifiedChannelSteeringVector}
	 [\boldsymbol{\alpha}(\boldsymbol{x},r,\theta)]_n  \approx e^{j2\pi\left(bx_n^2 - \Theta x_n\right)/\lambda}.
\end{align}
From \eqref{SimplifiedChannelSteeringVector}, the simplified channel steering vector is a function of $\boldsymbol{x}$, $b$, and $\Theta$. We denote the simplified channel steering vector as $\boldsymbol{\gamma}(\boldsymbol{x},b,\Theta)$ with 
\begin{align}\label{SimplifiedChannelSteeringVector2}
 [\boldsymbol{\gamma}(\boldsymbol{x},b,\Theta)]_n = e^{j2\pi\left(bx_n^2 - \Theta x_n\right)/\lambda}.
\end{align}

\begin{figure*}[!t]
	\centering
	\includegraphics[width=180mm]{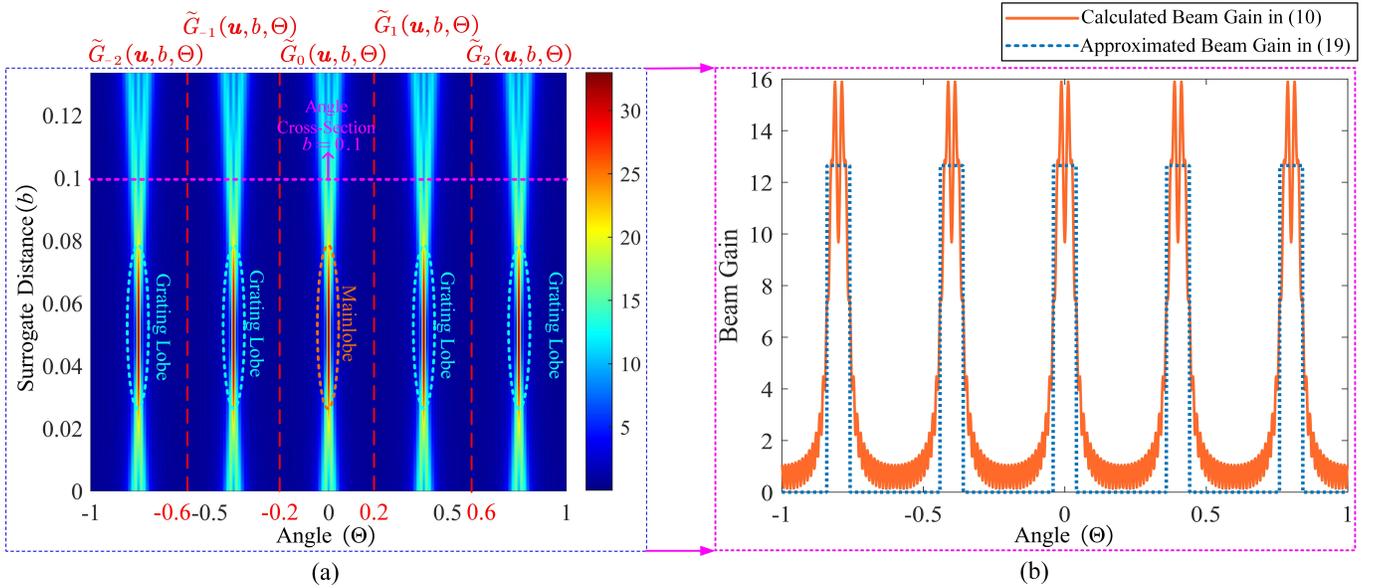}
	\caption{Illustration of $|G(\boldsymbol{u},b,\Theta)|$: (a) The calculated beam gain of $\boldsymbol{u}$ in the SD-A domain; (b) Comparisons of calculated and approximated beam gains.}
	\label{USLA5}
\end{figure*}

\section{Analysis of Uniform Sparse Arrays}\label{MCUSLA}
A direct approach to designing an SA is uniformly enlarging the antenna spacings of the HULA, which leads to the USA. First, we analyze the beam gains of channel steering vectors for USAs. Based on the analysis, we highlight three unique properties of the USAs, which shows that enlarging the antenna spacings  can effectively enhance the spatial resolution of the antenna arrays and improve the sum rate of multiuser communications.


In the context of USAs, the antennas are arranged uniformly in the antenna panel. Then, the x-axis coordinate of the $n$th antenna can be expressed as $\overline{x}_n = pn\lambda/2$. Substituting $\overline{x}_n$ into \eqref{SimplifiedChannelSteeringVector2}, the channel steering vector can be rewritten as
\begin{align}\label{USLAChannelSteeringVector1}
	[\boldsymbol{\gamma}(\boldsymbol{\overline{x}},b,\Theta)]_n = e^ {j\pi\left(\widetilde{b}p^2  n^2 - \Theta p  n\right)},
\end{align}
where $\boldsymbol{\overline{x}}\triangleq [\overline{x}_{-M},\cdots,\overline{x}_0,\cdots,\overline{x}_M]^{\rm T}$ and $\widetilde{b} \triangleq b\lambda/2$.  For an arbitrary channel steering vector $\boldsymbol{u}\triangleq\boldsymbol{\gamma}(\boldsymbol{\overline{x}},k,\Omega)$, its beam gain can be defined as 
\begin{align}\label{BeamGain}
	G(\boldsymbol{u},b,\Theta) &\triangleq N\boldsymbol{\gamma}(\boldsymbol{\overline{x}},b,\Theta)^{\rm H}\boldsymbol{u} \nonumber \\
	&= \sum_{n=-M}^{M}e^{j\pi\left(p(\Theta-\Omega)n - p^2\left(\widetilde{b}-\widetilde{k}\right)n^2\right)}\nonumber \\
	&\overset{\rm (a)}{=} \sum_{n=-M}^{M}e^{j\pi\left(p\left(\widetilde{\Theta}-\Omega\right)n - p^2\left(\widetilde{b}-\widetilde{k}\right)n^2\right)},
\end{align} 
where $\widetilde{k} \triangleq k\lambda/2$ and
\begin{align}\label{WidetildeOmega}
	\widetilde{\Theta} \triangleq \mbox{mod}\left(\Theta-\Omega+\frac{1}{p},\frac{2}{p}\right)-\frac{1}{p} + \Omega.
\end{align}
In \eqref{BeamGain}, $\rm (a)$ holds because of the periodicity of the complex sinusoidal functions. Following \cite{TWC23WZD}, we approximate the summation in \eqref{BeamGain} with the integral and have 
\begin{align}\label{BeamGain2}
	G(\boldsymbol{u},b,\Theta) & \approx \int_{-M-1/2}^{M+1/2} e^{j\pi\left(p\left(\widetilde{\Theta}-\Omega\right)z - p^2\left(\widetilde{b}-\widetilde{k}\right)z^2\right)}{\rm d}z\nonumber \\
	& = \int_{-\infty}^{\infty} U(z) e^{jJ(b,\Theta,z)}{\rm d}z,
\end{align}
where
\begin{align}\label{Uz}
	U(z) = \left\{ \begin{array}{ll}
		1, & -M-1/2\leq z \leq M+1/2,\\
		0, & \mbox{others},
	\end{array} \right.
\end{align}
and
\begin{align}\label{Jz}
J(b,\Theta,z) \triangleq \pi\big(p\big(\widetilde{\Theta}-\Omega\big)z - p^2\big(\widetilde{b}-\widetilde{k}\big)z^2\big).
\end{align}
An effective way to approximate the integral in \eqref{BeamGain2} is the principle of stationary phase (PSP)~\cite{AMMSE}, \cite{TWC24CKJ}. This method first determines the stationary phases of $J(b,\Theta,z)$ by finding $z$ that satisfies $J'(b,\Theta,z) = 0$. According to the expression of $J(b,\Theta,z)$, the zero points of $J'(b,\Theta,z)$ are 
\begin{align}\label{ZeroPoints}
	\widetilde{z}_m = \frac{\Theta-\Omega -2m/p}{2p\big(\widetilde{b}-\widetilde{k}\big)},
\end{align}
for $m\in\mathcal{T}$, where 
\begin{align}\label{SetS}
\mathcal{T}\triangleq\{m|m\in\mathbb{Z},|\Omega+2m/p|< 1 + 1/p\}.
\end{align}
 Then, based on the stationary phases, the PSP approximates  \eqref{BeamGain2} as 
\begin{align}\label{beamgainu}
	G(\boldsymbol{u},b,\Theta) &\approx \sum_{m\in\mathcal{T}}\sqrt{\frac{-2\pi}{J''(\widetilde{z}_m,b,\Theta)}}e^{-j\pi/4}U(\widetilde{z}_m)e^{jJ(b,\Theta,\widetilde{z}_m)}\nonumber \\
		&=\sum_{m\in\mathcal{T}} \frac{e^{-j\pi/4}}{\sqrt{p^2\big(\widetilde{b}-\widetilde{k}\big)}}U(\widetilde{z}_m)e^{jJ(b,\Theta,\widetilde{z}_m)} \nonumber\\
		& = \sum_{m\in\mathcal{T}}\widetilde{G}_m(\boldsymbol{u},b,\Theta),
\end{align}
where
\begin{align}\label{sub}
\widetilde{G}_m(\boldsymbol{u},b,\Theta) \triangleq \frac{e^{-j\pi/4}}{\sqrt{p^2\big(\widetilde{b}-\widetilde{k}\big)}}U(\widetilde{z}_m)e^{jJ(b,\Theta,\widetilde{z}_m)}.
\end{align}
The amplitudes of beam gains usually play a more significant role than the phases in the analysis of multiuser communications. Taking the absolute value of $\widetilde{G}_m(\boldsymbol{u},b,\Theta)$, we have
\begin{align}\label{SumSub}
	\big|\widetilde{G}_m(\boldsymbol{u},b,\Theta)\big| &= \frac{1}{\sqrt{p^2\big|\widetilde{b}-\widetilde{k}\big|}}U(\widetilde{z}_m) \nonumber \\
	&=\left\{ \begin{array}{ll}
		\frac{1}{\sqrt{p^2\big|\widetilde{b}-\widetilde{k}\big|}}, & -\frac{N}{2}\leq \widetilde{z}_m \leq \frac{N}{2}\\
		0, & \mbox{others}
	\end{array} \right. \nonumber \\
    &\overset{\rm (a)}{=} \left\{ \begin{array}{ll}
    	\frac{1}{\sqrt{p^2\big|\widetilde{b}-\widetilde{k}\big|}}, & \Theta\in\mathcal{B}_m\\
    	0, & \mbox{others},
    \end{array} \right.
\end{align}
where
\begin{align}\label{beamcoverage}
	\mathcal{B}_m\!\triangleq\!\left[\Omega\!-\!p\big|\widetilde{b}-\widetilde{k}\big|N\!+\!\frac{2m}{p},\Omega\!+\!p\big|\widetilde{b}-\widetilde{k}\big|N\!+\!\frac{2m}{p}\right].
\end{align}
In \eqref{SumSub}, we obtain $(\rm a)$ by considering the expression of  $\widetilde{z}_m$  in \eqref{ZeroPoints}. 


Fig.~\ref{USLA5} illustrates the absolute beam gain of $\boldsymbol{u}$. For the USA, we set $N=33$, $\lambda = 0.01$~m and  $p = 5$. For the channel steering vector $\boldsymbol{u}$, we set $k = 0.05$ and $\Omega = 0$, which corresponds to $r = 10$~m. In Fig.~\ref{USLA5}(a),  we illustrate the calculated beam gain of $\boldsymbol{u}$. The y-axis and the x-axis represent the surrogate distance ($b$) and angle ($\Theta$), respectively. Therefore, we term the spatial domain in Fig.~\ref{USLA5}(a) as the surrogate distance-angle (SD-A) domain, where the surrogate distance and angle serve as the primary coordinates for characterizing the channel steering vectors. In Fig.~\ref{USLA5}(b), we compare the calculated beam gain in \eqref{BeamGain} with the approximated beam gain in \eqref{SumSub} by taking the angle cross-section as an example, where we set $b = 0.1$. From the figure, the concise approximation in \eqref{SumSub}  accurately characterizes the spatial region and beam gain of $\boldsymbol{u}$ in the SD-A domain, which provides an intuitive understanding of the correlations between near-field channel steering vectors. Based on the analysis from \eqref{BeamGain} to \eqref{beamcoverage} and the intuitive illustration in Fig.~\ref{USLA5}, we then delve into exploring the unique properties of channel steering vectors for USAs. First, we summarize the overall characteristics in \textbf{Property~1}.

\textbf{Property 1 (Overall Characteristics):} The beam pattern of $\boldsymbol{u}$ is periodic with a period of $2/p$, where the beam gain and beam coverages in each period are approximated in \eqref{SumSub} and \eqref{beamcoverage}, respectively.

This property can be  analytically verified based on the analysis from \eqref{BeamGain} to \eqref{beamcoverage} and intuitively checked based on the illustration in Fig.~\ref{USLA5}.   From the figure, the beam gain of $\boldsymbol{u}$ is constituted of $p=5$ parts and the period is $2/p =0.4$, which are consistent with \eqref{SetS} and \eqref{beamgainu}. In addition, the beam patterns of different parts are the same, following the approximations in \eqref{SumSub} and~\eqref{beamcoverage}.

\begin{figure}[!t]
	\centering
	\includegraphics[width=75mm]{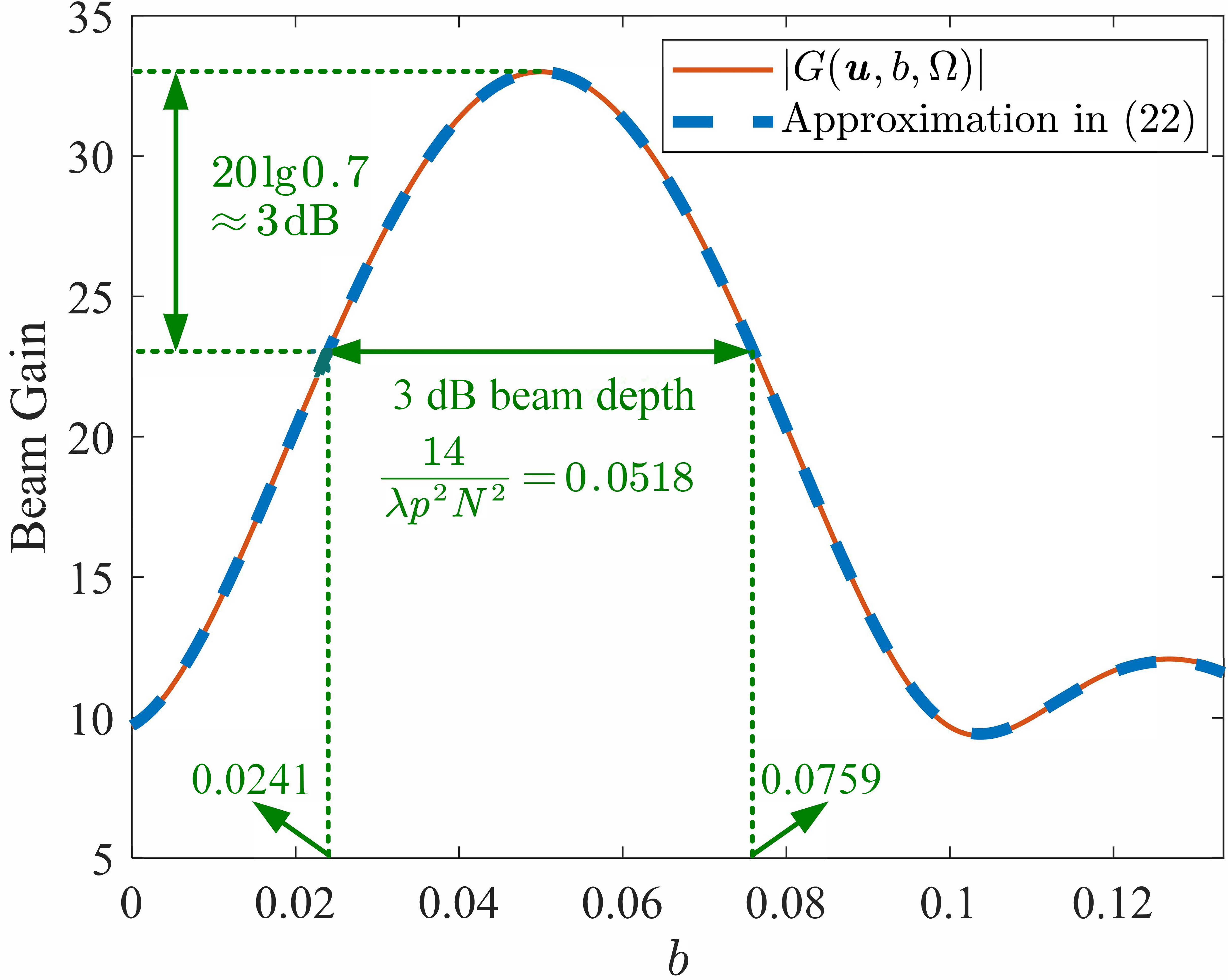}
	\caption{Illustration of $|G(\boldsymbol{u},b,\Omega)|$.}
	\label{DistCrossSection}
\end{figure}

Due to the relatively small array aperture, the beamforming of the conventional HULA can only steer energy to a specific direction. However, as observed in Fig.~\ref{USLA5}, the beamforming of USAs allows for precise energy focusing on a specific region. To systematically assess this beam focusing ability in the SD-A domain, we introduce \textbf{Property 2}. 

\textbf{Property 2 (Beam Focusing):} For the beam gain of $\boldsymbol{u}$, we denote the beamwidth and beam depth of its mainlobe as $B_{\rm USA}$ and $\widetilde{B}_{\rm USA}$, respectively. Then, we have  $B_{\rm USA}= \frac{2}{pN}$ and $\widetilde{B}_{\rm USA}= \min\left\{\frac{14}{\lambda p^2N^2},b_{\rm max}\right\}$.

 \textit{Proof:} To compute $B_{\rm USA}$, we first resort to the angle cross-section of the mainlobe, which is $|G(\boldsymbol{u},k,\Theta)|$. From \eqref{BeamGain}, we have
 \begin{align}\label{SparseArray}
 	|G(\boldsymbol{u},k,\Theta)| &= \left|\sum_{n=-M}^{M}e^{j\pi p(\Theta-\Omega)n}\right| \nonumber\\
 	&= \left|\frac{\sin(\pi p (\Theta-\Omega)N/2)}{\sin(\pi (\Theta-\Omega)N/2)}\right|.
 \end{align}
In fact, \eqref{SparseArray} represents the array response of a far-field channel steering vector for USAs and its beamwidth is typically set as $\frac{2}{pN}$. Therefore, we have $B_{\rm USA}= \frac{2}{pN}$. 

To compute $\widetilde{B}_{\rm USA}$, we  resort to the distance cross-section of the mainlobe, which is $|G(\boldsymbol{u},b,\Omega)|$. From \eqref{BeamGain}, we have
 \begin{align}\label{SparseArray2}
	|G(\boldsymbol{u},b,\Omega)| &= \left| \sum_{n=-M}^{M}e^{j\pi(p^2(\widetilde{b}-\widetilde{k})n^2)}\right| \nonumber \\
	&  \overset{\rm (a)}{\approx}\! \left|\int_{-M-1/2}^{M+1/2} e^{j\pi(p^2(\widetilde{b}-\widetilde{k})z^2)}{\rm d}z\right| \nonumber \\
	& = \sqrt{\frac{2\mathcal{C}\left(\zeta\right)^2 + 2\mathcal{S}\left(\zeta\right)^2}{p^2|\widetilde{b}-\widetilde{k}|}},
\end{align}
where we approximate the summation as integral in $\rm (a)$, $\zeta\triangleq\sqrt{2p^2|\widetilde{b}\!-\!\widetilde{k}|}(M+1/2)$, $\mathcal{C}(\zeta) \triangleq \int_0^{\zeta} \cos(\pi z^2/2){\rm d} z$ and $\mathcal{S}(\zeta) \triangleq \int_0^{\zeta} \sin(\pi z^2/2){\rm d} z$ are the Fresnel functions. As shown in Fig.~\ref{DistCrossSection}, we illustrate the beam gain of the distance cross-section $|G(\boldsymbol{u},b,\Omega)|$, where the parameter settings are the same as in Fig.~\ref{USLA5}. From the figure, the absolute beam gains, $|G(\boldsymbol{u},b,\Omega)|$, can be well approximated with  the expressions in \eqref{SparseArray2}. By substituting $|\widetilde{b}-\widetilde{k}| = \frac{\kappa}{p^2N^2}$, where $\kappa$ is the scaling factor, into \eqref{SparseArray2}, we have 
\begin{align}\label{NormalizedBeamGain}
	|G(\boldsymbol{u},b,\Omega)|/N = \sqrt{\frac{2\mathcal{C}(\sqrt{\kappa/2})^2 + 2\mathcal{S}(\sqrt{\kappa/2})^2}{\kappa}}. 
\end{align}
From \eqref{NormalizedBeamGain}, $|G(\boldsymbol{u},b,\Omega)|/N$ is only related to $\kappa$. Therefore, we can determine the beam depth for different numbers of antennas by selecting a suitable $\kappa$. When $\kappa = 3.5$, we have $|G(\boldsymbol{u},b,\Omega)|/N \approx 0.7036$, which corresponds to $-3.05$~dB. Note that $\widetilde{b} = b\lambda/2$ and $\widetilde{k} = k\lambda/2$. Therefore, the 3 dB beam depth is approximately $\frac{14}{\lambda p^2N^2}$. In some scenarios, where the array aperture is small, the 3~dB beam depth may be larger than the BS coverage. Therefore, we normalize $\widetilde{B}_{\rm USA}$~as 
 \begin{align}\label{USLADistance}
\widetilde{B}_{\rm USA} = \min\left\{\frac{14}{\lambda p^2N^2},b_{\rm max}\right\},
\end{align}
which completes the proof.

\textbf{Property 2} indicates that the beamforming of the USAs has the ability to focus energy on a specific region in near field. In addition, the beamwidth and beam depth of the mainlobe are related to both the number of antennas, $N$, and the array sparsity factor, $p$. Specifically, the beamwidth decreases linearly with $pN$ while the beam depth decreases quadratically with $pN$.  Therefore, besides increasing the number of antennas, increasing the antenna spacing to form USAs can also improve the beam focusing ability and therefore enhance the spatial resolution. Note that the evaluation of near-field beam focusing ability has been conducted in previous works, such as \cite{OJCS23LYW} and \cite{ACSSC21BE}. However, these works assess beam focusing in the physical space, where the beam focusing ability is nonuniform and dependent on the specific locations of the focused points. In contrast, in \textbf{Property 2}, the evaluation of beam focusing is performed for the considered USA in the SD-A domain, where the beam focusing ability remains uniform across all locations and is solely determined by the array configurations. The concise expression in \textbf{Property 2} gives a more intuitive understanding of the near-field beam focusing than the existing works.

From \textbf{Property 1},  the beam gain of $\boldsymbol{u}$ has several mainlobes and those that do  not fit with the channel parameters are usually termed as the grating lobes. In multiuser communications, the users distribute randomly in the space. When the users are located in the grating lobe, the strong interference between users may significantly deteriorate the sum rate of multiuser communications.   To dispel this concern, we introduce \textbf{Property 3}.

\textbf{Property 3 (Spatial Resolution):} Denote the total coverage of the mainlobes for a  USA as $C_{\rm USA}$. Denote the coverage of the mainlobe for a conventional  HULA as $C_{\rm HULA}$. Then we have $C_{\rm USA} \le C_{\rm HULA}$.

\textit{Proof:}  Since the HULA usually has a small array aperture, the beam depth of the its mainlobe in the SD-A domain is  the BS coverage, i.e., $\widetilde{B}_{\rm HULA} = b_{\rm max}$. In addition, the beamwidth of the mainlobe of the HULA is typically set as $B_{\rm HULA} = 2/N$. Therefore, we have 
\begin{align}\label{CHULA}
C_{\rm HULA} = B_{\rm HULA}\widetilde{B}_{\rm HULA} = 2b_{\rm max}/N.
\end{align}
On the other hand, the USA has $p$ mainlobes and the coverage of each mainlobe can be expressed as $B_{\rm USA}\widetilde{B}_{\rm USA}$. With \textbf{Property 2}, we have 
\begin{align}\label{CUSLA}
C_{\rm USA} = pB_{\rm USA}\widetilde{B}_{\rm USA}\le 2b_{\rm max}/N.
\end{align}
Comparing \eqref{CHULA} and \eqref{CUSLA}, we have $C_{\rm USA} \le C_{\rm HULA}$, which completes the proof.

Although the beamforming of the USA usually has grating lobes, the union of the grating lobes still has a smaller coverage than the conventional HULA according to \textbf{Property~3}. In other words, the USA can provide a higher spatial resolution than the HULA  for multiuser communications even under the influence of grating lobes. This improved spatial resolution can lead to the higher sum rate of multiuser communications with the  USA  than that with  the HULA. Note that the angle resolution of USAs has been analyzed in~\cite{ACSSC21BE}. However, it does not address the comparison of spatial resolution, which encompasses both the angle resolution and distance resolution, between USAs and HULAs.

\section{Antenna Position Optimization of Nonuniform Sparse Arrays}\label{MUNSLA}
One potential challenge associated with the USAs is the grating lobes induced by the identical and wide spacing between antennas.  To address this challenge, we propose the NSAs.  By introducing additional DoF through adjusting antenna spacings, NSAs  offer a  viable solution to mitigating grating lobes and enhancing system performance. To maximize the sum rate of near-field multiuser communications, we optimize the antenna positions of the NSAs. Since the antenna position optimization problem is nonconvex, an SCA-APO algorithm is proposed. 

The multiuser sum rate can be expressed as 
\begin{align}
	R_{\rm sum} = \sum_{k=1}^K \log_2(1+\Gamma_k).
\end{align} 
$\Gamma_k$ denotes the signal-to-interference-plus-noise ratio of the $k$th user and  can be expressed as 
\begin{align}
	\Gamma_k = \frac{|\boldsymbol{h}_k^{\rm H}\boldsymbol{f}_k|^2}{\sum_{i=1,i\neq k}^{K}|\boldsymbol{h}_k^{\rm H}\boldsymbol{f}_i|^2 + \sigma^2},
\end{align}
where $\boldsymbol{f}_k$ denotes the beamformer for the $k$th user. In practice, once antennas are installed, their positions are typically fixed and cannot be easily changed. Consequently, we need to optimize the antenna positions by considering all potential multiuser channels rather than focusing solely on the channels observed in specific instances. Then the antenna position optimization problem can be formulated  as 
\begin{subequations}\label{SumRateProblem1} \normalsize
\begin{align}
\mbox{(P1)}~~~~~~~\max_{\boldsymbol{x}}~&\mathbb{E}\{R_{\rm sum}\} \label{SumRateObjective1}\\
	{\rm s.t.}~~~&|x_n-x_m|\ge\lambda/2 \label{SumRateConstraint1}\\
	 &x_n\ge-D/2,~x_n\le D/2 \label{SumRateConstraint2}\\
	 &m,n = -M,\cdots,0,\cdots M,~m\neq n, \nonumber 
\end{align}
\end{subequations}
where the objective in \eqref{SumRateObjective1} aims at maximizing the expectation of the sum rate with respect to $\boldsymbol{h}_k$,  the constraint in \eqref{SumRateConstraint1} denotes the minimum antenna spacing limitation to avoid the coupling effects between adjacent antennas, and the constraint in \eqref{SumRateConstraint2} denotes the space limitation of the antenna panel. Indeed, solving the problem in \eqref{SumRateProblem1} is difficult due to three challenges:  1) Calculating the expectation of the sum rate with respect to $\boldsymbol{h}_k$ is challenging due to the highly nonlinear relationships.  2) The minimum antenna spacing constraint in \eqref{SumRateConstraint1} is not  convex. 3) The objective in \eqref{SumRateObjective1} is a nonconvex function of the antenna positions. Subsequently, we  focus on overcoming these three challenges  and find solutions for \eqref{SumRateProblem1}.
\subsection{Overcoming the First Challenge}
According to Jensen's inequality, we have 
\begin{align}
	\mathbb{E}\{R_{\rm sum}\} \geq   \sum_{k=1}^K \log_2(1+  \big(\mathbb{E}\{\Gamma_k^{-1}\}\big)^{-1}).
\end{align}
To streamline our analysis, we opt to optimize the lower bound of $\mathbb{E}\{R_{\rm sum}\}$, i.e., $\sum_{k=1}^K \log_2(1+  \mathbb{E}\{\Gamma_k^{-1}\}^{-1})$. In addition, $\mathbb{E}\{\Gamma_k^{-1}\} $ would be the same for all the users since the expectation transverses all possibilities of the channels.  Therefore, we focus on the analysis of an arbitrary user and convert \eqref{SumRateProblem1} to 
\begin{subequations}\label{SumRateProblem2} \normalsize
	\begin{align}
\mbox{(P2)}~~~~~~~\min_{\boldsymbol{x}}~~&\mathbb{E}\{\Gamma_k^{-1}\}\label{SumRateObjective2}\\
		{\rm s.t.}~~~&\eqref{SumRateConstraint1}~\mathrm{and}~\eqref{SumRateConstraint2}.
	\end{align}
\end{subequations} 
To further simplify \eqref{SumRateProblem2}, we adopt the maximum ratio combining, which is widely employed for multiuser sum rate analysis due to its ability to assess the MUI~\cite{CL22LHQ}.  By setting $\boldsymbol{f}_k = \boldsymbol{h}_{k}/\boldsymbol{h}_{k}^{\rm H}\boldsymbol{h}_{k}$, for $k=1,2,\cdots,K$, the denominator of $\mathbb{E}\{\Gamma_k^{-1}\}$ will be a constant. Thus, the minimization of $\mathbb{E}\{\Gamma_k^{-1}\}$ in \eqref{SumRateObjective2} can be converted to the minimization of the numerator of $\mathbb{E}\{\Gamma_k^{-1}\}$, i.e., $\mathbb{E}\big\{\sum_{i=1,i\neq k}^{K}|\boldsymbol{h}_k^{\rm H}\boldsymbol{h}_i|^2/|\boldsymbol{h}_{i}^{\rm H}\boldsymbol{h}_{i}|^2 + \sigma^2\big\}$. Note that
\begin{align}\label{SumRateObjective3}
	&~~~~\mathbb{E}\left\{\sum_{i=1,i\neq k}^{K}\big|\boldsymbol{h}_k^{\rm H}\boldsymbol{h}_i\big|^2 /\big|\boldsymbol{h}_{i}^{\rm H}\boldsymbol{h}_{i}\big|^2 + \sigma^2\right\} \nonumber \\
	&\overset{\rm (a)}{=}\!\mathbb{E}\left\{\sum_{i=1,i\neq k}^{K}\big|\boldsymbol{h}_k^{\rm H}\boldsymbol{h}_i\big|^2 /\big|\boldsymbol{h}_{i}^{\rm H}\boldsymbol{h}_{i}\big|^2\right\}+ \sigma^2 \nonumber \\
	&\overset{\rm (b)}{=} \!(K-1)\mathbb{E}\left\{\big|\boldsymbol{h}_k^{\rm H}\boldsymbol{h}_i\big|^2 /\big|\boldsymbol{h}_{i}^{\rm H}\boldsymbol{h}_{i}\big|^2\right\}+ \sigma^2,
\end{align}
where $\rm (a)$ holds because $\sigma^2$ is a constant and $\rm (b)$ holds because the expectation of MUI is the same for all users. Then, the objective in \eqref{SumRateProblem2} can be converted to the minimization of 
\begin{align}\label{expectation22}
	&~~~~\mathbb{E}\left\{\big|\boldsymbol{h}_k^{\rm H}\boldsymbol{h}_i\big|^2/\big|\boldsymbol{h}_{i}^{\rm H}\boldsymbol{h}_{i}\big|^2\right\} \nonumber \\
&\overset{\rm (a)}{=}\sum_{l=1}^{L_k}\sum_{u=1}^{L_i} \mathbb{E}\left\{\!\frac{\left|\!\gamma_k^{(l)}\!\gamma_i^{(u)}\!\right|^2}{\big|\boldsymbol{h}_{i}^{\rm H}\boldsymbol{h}_{i}\big|^2}\!\left|\boldsymbol{\alpha}\left(\!\boldsymbol{x},\!r_k^{(l)},\!\theta_k^{(l)}\!\right)^{\rm H} \boldsymbol{\alpha}\left(\!\boldsymbol{x},\!r_i^{(u)},\!\theta_i^{(u)}\!\right)\right|^2\!\right\}\nonumber\\
&\overset{\rm (b)}{=}\sum_{l=1}^{L_k}\sum_{u=1}^{L_i}\xi_k^{(l)}\widetilde{\xi}_i^{(u)}\mathbb{E}\left\{\left|\boldsymbol{\alpha}\left(\boldsymbol{x},\!r_k^{(l)},\!\theta_k^{(l)}\!\right)^{\rm H} \boldsymbol{\alpha}\left(\!\boldsymbol{x},\!r_i^{(u)},\!\theta_i^{(u)}\!\right)\right|^2\right\},
\end{align}
where $\rm (a)$ holds because $\mathbb{E}\big\{\gamma_k^{(l)}\big\} = 0$, and $\rm (b)$ holds because $\xi_k^{(l)}=\mathbb{E}\big\{|\gamma_k^{(l)}|^2\big\}$ and $\widetilde{\xi}_i^{(u)} \triangleq\mathbb{E}\big\{\big|\!\gamma_i^{(u)} \big|^2/{\big|\boldsymbol{h}_{i}^{\rm H}\boldsymbol{h}_{i}\big|^2}\big\}$. Due to the expectation operation, the minimization of \eqref{expectation22} can be converted to the minimization of 
\begin{align}\label{expectation23}
&~~~~\mathbb{E}\left\{\left|\boldsymbol{\alpha}\left(\boldsymbol{x},\!r_k^{(l)},\!\theta_k^{(l)}\!\right)^{\rm H} \boldsymbol{\alpha}\left(\!\boldsymbol{x},\!r_i^{(u)},\!\theta_i^{(u)}\!\right)\right|^2\right\} \nonumber \\
&\overset{\rm (a)}{\approx}\!\mathbb{E}\left\{\left|\boldsymbol{\gamma}\left(\boldsymbol{x},b_k^{(l)},\Theta_k^{(l)}\right)^{\rm H}\boldsymbol{\gamma}\left(\boldsymbol{x},b_i^{(u)},\Theta_i^{(u)}\right)\right|^2\right\} \nonumber \\
& =\mathbb{E}\left\{\!\left|\! \sum_{n=-M}^{M}\!e^{j2\pi\left(\left(b_i^{(u)} - b_k^{(l)}\right)x_n^2 + \left(\Theta_k^{(l)} - \Theta_i^{(u)}\right)x_n\right)/\lambda}\! \right|^2 \!\right\} \nonumber \\
& \overset{\rm (b)}{=}\!  \mathbb{E}\left\{\!\left| \sum_{n=-M}^{M}\!e^{j2\pi\big(\overline{\overline{b}}x_n^2 + \overline{\overline{\Theta}}x_n\big)/\lambda}\!\right|^2 \!\right\},
\end{align}
where  $\rm (a)$ holds by adopting the approximation in \eqref{SimplifiedChannelSteeringVector2}, and we define $\overline{\overline{b}}\triangleq b_i^{(u)} - b_k^{(l)}$ and $\overline{\overline{\Theta}}\triangleq \Theta_k^{(l)} - \Theta_i^{(u)}$ in $\rm (b)$. Note that \eqref{expectation23} essentially represents the correlations between  channel steering vectors, which  depend solely on the antenna configurations and are irrelevant to the specific channel state information.

One remaining problem is how to calculate  the expectation in \eqref{expectation23}. Obtaining the closed-form solution of the expectation in \eqref{expectation23} is challenging due to the highly nonlinear relationships. Therefore, we turn to calculating \eqref{expectation23} via numerical methods. Note that $b_i^{(u)},b_k^{(l)}\in[0,b_{\rm max}]$ and $\Theta_k^{(l)}, \Theta_i^{(u)}\in[-1,1]$. Therefore, we have $\overline{\overline{b}}\in[-b_{\rm max},b_{\rm max}]$ and $\overline{\overline{\Theta}} \in[-2,2]$. We quantize the intervals of $\overline{\overline{b}}$ and $\overline{\overline{\Theta}}$ into $S$ and $T$ samples, respectively, where the $s$th sample of  $\overline{\overline{b}}$ and $t$th sample of $\overline{\overline{\Theta}}$  can be expressed as 
\begin{align}\label{BeamSamples}
&\widehat{b}_s = -b_{\rm max} + \frac{2(s-1)b_{\rm max}}{S}, \nonumber \\
 &\widehat{\Theta}_t = -2 + \frac{4(t-1)}{T}.
\end{align}
Then, we calculate the distribution of different samples. Without loss of generality, we assume $b_i^{(u)}$ and $b_k^{(l)}$ follow the uniform distribution within $[0,b_{\rm max}]$ while $\Theta_k^{(l)}$ and $\Theta_i^{(u)}$ follow the uniform distribution within $[-1,1]$. \footnote{In fact, we only need to change \eqref{bdistribution} and \eqref{Thetadistribution} to adapt to other distributions, which will not change the following procedures.} The probability distribution functions of $\overline{\overline{b}}$ and $\overline{\overline{\Theta}}$ can be expressed~as 
\begin{align}\label{bdistribution}
	f\left(\overline{\overline{b}}\right) = \frac{1}{b_{\rm max}} - \frac{1}{b_{\rm max}^2} \left|\overline{\overline{b}}\right|,~\mbox{for}~\overline{\overline{b}}\in[-b_{\rm max},b_{\rm max}],
\end{align}
and
\begin{align}\label{Thetadistribution}
		g\left(\overline{\overline{\Theta}}\right) = \frac{1}{2} - \frac{1}{4}\left|\overline{\overline{\Theta}}\right|,~\mbox{for}~\overline{\overline{\Theta}}\in[-2,2].
\end{align}
With \eqref{BeamSamples}, \eqref{bdistribution} and \eqref{Thetadistribution},  the expectation in \eqref{expectation23} can be expressed as
\begin{align}\label{FinalExpectation}
	&~~~~\mathbb{E}\left\{\!\left| \sum_{n=-M}^{M}\!e^{j2\pi\big(\overline{\overline{b}}x_n^2 + \overline{\overline{\Theta}}x_n\big)/\lambda} \right|^2 \!\right\} \nonumber \\
	& \approx\!\frac{1}{ST}\sum_{s=1}^{S}\sum_{t=1}^{T} w_{t,s}\left| \sum_{n=-M}^{M}\!e^{j2\pi\big(\widehat{b}_s x_n^2 +\widehat{\Theta}_t x_n\big)/\lambda} \right|^2\nonumber \\
	&\triangleq\! h(\boldsymbol{x}),
\end{align}
where $w_{t,s}\triangleq f\big(\widehat{b}_s\big)g\big(\widehat{\Theta}_t\big)$. 
\subsection{Overcoming the Second Challenge}
Now, we turn to solving the nonconvex constraint in \eqref{SumRateConstraint1}. This constraint is designed to regulate the separation between adjacent antennas. In the context of linear arrays, the antennas are arranged in a line. We can ensure adherence to the minimum spacing constraint by maintaining a specified distance between the current antenna and the former one. Consequently, the constraint in \eqref{SumRateConstraint1} can be converted to
\begin{align}\label{antennaspace}
	x_n - x_{n-1} > \lambda/2,~\mbox{for}~n = -M+1,\cdots,0,\cdots,M,
\end{align}
which is a convex constraint. 

\subsection{Overcoming the Third Challenge}
Based on \eqref{FinalExpectation} and \eqref{antennaspace}, (P2) can be converted to 
\begin{subequations}\label{SumRateProblem3} \normalsize
	\begin{align}
		\mbox{(P3)}~~~~~~~\min_{\boldsymbol{x}}~&h(\boldsymbol{x}) \label{SumRateObjective4}\\
		{\rm s.t.}~~&\eqref{antennaspace}~\mathrm{and}~\eqref{SumRateConstraint2}.~~~~~~~~~~~
	\end{align}
\end{subequations}
Note that (P3) is an optimization problem with convex constraints but nonconvex objective. To deal with the nonconvex objective in \eqref{SumRateObjective4}, we then propose a successive convex approximation-based antenna position optimization (SCA-APO) algorithm.

First of all, we randomly initialize the antenna  positions as~$\boldsymbol{x}^{(0)}$.

In the $q$th iteration, for $q\ge1$, the key point of the SCA is to find a convex surrogate function that can locally approximate the original function around $\boldsymbol{x}^{(q-1)}$ and is also the upper bound of the original function. According to the Taylor's theorem~\cite{Magnus1995}, we have
\begin{align}\label{key}
	h(\boldsymbol{x})&\leq h\big(\boldsymbol{x}^{(q-1)}\big) + \nabla h\big(\boldsymbol{x}^{(q-1)}\big)^{\rm T}\big(\boldsymbol{x} -\boldsymbol{x}^{(q-1)}\big) \nonumber \\
	&+ \frac{\chi}{2}\big(\boldsymbol{x} -\boldsymbol{x}^{(q-1)}\big)^{\rm T}\big(\boldsymbol{x} -\boldsymbol{x}^{(q-1)}\big)\nonumber \\
	&\triangleq \overline{h}\big(\boldsymbol{x},\boldsymbol{x}^{(q-1)}\big),
\end{align}
where $\nabla h\big(\boldsymbol{x}^{(q-1)}\big)$ denotes the gradient vector and $\chi$ denotes the maximum eigenvalue of the Hessian matrix $\nabla^2 h\big(\boldsymbol{x}^{(q-1)}\big)$. The gradient vector $\nabla h\big(\boldsymbol{x}\big)$  and the Hessian matrix $\nabla^2 h\big(\boldsymbol{x}\big)$ are provided in \eqref{gradientvector} and \eqref{Hessianmatrix}, respectively, which are shown at the top of the next page. 
\begin{figure*}[htbp]
	\begin{align}\label{gradientvector}
		\big[\nabla h\big(\boldsymbol{x}\big)\big]_n = \sum_{s=1}^{S}\sum_{t=1}^{T}\frac{4\pi w_{t,s}(2\widehat{b}_sx_n +\widehat{\Theta}_t)}{ST\lambda}\sum_{v=-M}^{M}\sin\left(\frac{2\pi}{\lambda}\big(\widehat{b}_s(x_v^2-x_n^2) + \widehat{\Theta}_t(x_v-x_n)\big)\right).
	\end{align}
	\hrulefill
\end{figure*}
\begin{figure*}[htbp]
	\begin{align}\label{Hessianmatrix}
		\big[\nabla^2h\big(\boldsymbol{x}\big)\big]_{m,n} = \left\{ \begin{array}{ll}
			\sum_{s=1}^{S}\sum_{t=1}^{T}\frac{8\pi^2 w_{t,s}(2\widehat{b}_sx_n +\widehat{\Theta}_t)(2\widehat{b}_sx_m +\widehat{\Theta}_t)}{ST\lambda^2} \cos\left(\frac{2\pi}{\lambda}\big(\widehat{b}_s(x_m^2-x_n^2) + \widehat{\Theta}_t(x_m-x_n)\big)\right), & m\neq n,\\
			\sum_{s=1}^{S}\sum_{t=1}^{T}\frac{8\pi w_{t,s} \widehat{b}_s}{ST\lambda}\sum_{v=-M}^{M}\sin\left(\frac{2\pi}{\lambda}\big(\widehat{b}_s(x_v^2-x_n^2) + \widehat{\Theta}_t(x_v-x_n)\big)\right), & m=n.
		\end{array} \right.
	\end{align}
	\hrulefill
\end{figure*}
Then, the $q$th subproblem can be expressed as
\begin{subequations}\label{SumRateProblem4} \normalsize
	\begin{align}
		(\mbox{P3-}q)~~~~~~~~\min_{\boldsymbol{x}}~&\overline{h}\big(\boldsymbol{x},\boldsymbol{x}^{(q-1)}\big) \label{SumRateObjective5}\\
		{\rm s.t.}~~&\eqref{antennaspace}~\mathrm{and}~\eqref{SumRateConstraint2}.~~~~~~~~~~~
	\end{align}
\end{subequations}
Obviously, $(\mbox{P3-}q)$ is a convex problem and therefore can be effectively solved. We omit the details and denote the solution of $(\mbox{P3-}q)$ as $\boldsymbol{x}^{(q)}$.

We iteratively solve $(\mbox{P3-}q)$ until the maximum number of iterations $Q$ is reached, where we denote  the optimized antenna position as $\widehat{\boldsymbol{x}}$. Finally, we summarize the SCA-APO algorithm in \textbf{Algorithm~\ref{SCA-APO}}.

\begin{algorithm}[!t]
	\caption{Successive Convex Approximation-based Antenna Position Optimization (SCA-APO) Algorithm}
	\label{SCA-APO}
	\begin{algorithmic}[1]
		\STATE \textbf{Input:} $D$, $Q$ and $N$.		
		\STATE \textbf{Initialization:} $q\leftarrow 0$. Generate $\boldsymbol{x}^{(0)}$ randomly.
		\WHILE{$q < Q$}
		\STATE 	$q\leftarrow q + 1 $.
		\STATE  Obtain $\boldsymbol{x}^{(q)}$ by solving $(\mbox{P3-}q)$.
		\ENDWHILE  
		\STATE $\widehat{\boldsymbol{x}}\leftarrow \boldsymbol{x}^{(Q)}$.
		\STATE \textbf{Output:}  $\widehat{\boldsymbol{x}}$.
	\end{algorithmic}
\end{algorithm}

Now we analyze the convergence of the SCA-APO algorithm. Note that 
\begin{align}\label{convergence}
h(\boldsymbol{x}^{(q)})\!\overset{\rm (a)}{\le}\!\overline{h}(\boldsymbol{x}^{(q)},\boldsymbol{x}^{(q-1)})\!\overset{\rm  (b)}{\le}\! \overline{h}(\boldsymbol{x}^{(q-1)},\boldsymbol{x}^{(q-1)})\!\overset{\rm (c)}{=}\!h(\boldsymbol{x}^{(q-1)}),
\end{align}
where we obtain $(\rm a)$ according to \eqref{key}, obtain $\rm  (b)$ because $\boldsymbol{x}^{(q)}$ is the optimal solution of \eqref{SumRateProblem4}, and obtain $\rm  (c)$ by comparing the expressions of $\overline{h}\big(\boldsymbol{x}^{(q-1)},\boldsymbol{x}^{(q-1)}\big)$ and $h\big(\boldsymbol{x}^{(q-1)}\big)$ in \eqref{key}. According to \eqref{convergence}, we have $h\big(\boldsymbol{x}^{(q)}\big)\le h\big(\boldsymbol{x}^{(q-1)}\big)$, which indicates that the objective value will decrease with the iteration and therefore the SCA-APO algorithm converges.

We also analyze the computational complexity of the SCA-APO algorithm. The SCA-APO algorithm contains $Q$ iterations. In each iteration, the computational complexity mainly comes from solving \eqref{SumRateProblem4}. In fact, \eqref{SumRateProblem4} is a typical convex quadratic programming problem and its computational complexity is $\mathcal{O}(N^{3.5})$. In total, the computational complexity of the proposed SCA-APO algorithm is $\mathcal{O}(QN^{3.5})$.

%
%

\section{Channel Estimation and Precoding}\label{CEP}
In this section, we first explore the channel sparsity for USAs and NSAs. Based on the sparsity in the SD-A domain, an on-grid SDA-OMP algorithm is proposed to estimate multiuser channels for SAs. To further improve the resolution of the SDA-OMP, an off-grid SDA-ISRCE algorithm is further developed. Then,  beamforming is performed to mitigate the MUI based on the estimated channels.

\subsection{Sparsity Exploration for SAs}
\subsubsection{Sparsity Exploration for USAs}From  \textbf{Property 2},  the beamforming of USAs allows for energy focusing on a specific point in the SD-A domain. Therefore, the channels of USAs exhibit sparsity in the SD-A domain. However, \textbf{Property 2} only focuses on a specific channel steering vector.  To unveil the relationship among multiple channel steering vectors, we have  \textbf{Proposition 1}.

\textbf{Proposition 1 (Translation Invariance):} For another channel steering vector, $\overline{\boldsymbol{u}} = \boldsymbol{\gamma}(\boldsymbol{\overline{x}},\overline{k},\overline{\Omega})$, we have  $G(\overline{\boldsymbol{u}},b,\Theta) = G(\boldsymbol{u},b + \Delta k,\Theta + \Delta \Omega)$, where $\Delta k = k - \overline{k}$ and $\Delta \Omega = \Omega - \overline{\Omega}$.

 \textit{Proof:} Define $\widehat{k} \triangleq \overline{k}\lambda/2$. Then, from \eqref{BeamGain}, we have
\begin{align}\label{BeamGain3}
	G(\overline{\boldsymbol{u}},b,\Theta) &= N\boldsymbol{\gamma}(\boldsymbol{\overline{x}},b,\Theta)^{\rm H}\overline{\boldsymbol{u}}\nonumber \\
	& = \sum_{n=-M}^{M}e^{j\pi\left(p\left(\Theta  - \overline{\Omega}\right)n - p^2\left(\widetilde{b}-\widehat{k}\right)n^2\right)} \nonumber \\
	& = \sum_{n=-M}^{M}e^{j\pi\left(p\left(\Theta-\Omega + \Omega - \overline{\Omega}\right)n - p^2\left(\widetilde{b}-\widetilde{k} + \widetilde{k} - \widehat{k}\right)n^2\right)} \nonumber \\
	&= G\left(\boldsymbol{u},b + k - \overline{k},\Theta + \Omega - \overline{\Omega}\right) \nonumber \\
	& =  G\left(\boldsymbol{u},b + \Delta k,\Theta + \Delta \Omega\right),
\end{align}
which completes the proof. 

\textbf{Proposition 1} indicates that the beam gain of $\overline{\boldsymbol{u}}$ is the translation of that of $\boldsymbol{u}$.  As a result, different from the polar-domain sparsity in~\cite{Tcom22CMH},  the SD-A-domain sparsity of USA channels is uniform. Moreover, from \textbf{Property 1}, the focused points  are periodic in the SD-A domain  with a period  of $2/p$. Therefore, the channels of the USA also exhibit periodic sparsity in the SD-A domain besides the uniform sparsity. Based on this sparsity,  we establish an SD-A-domain representation of the USA channels by uniformly sampling the surrogate distance $b$ within $[0,b_{\rm max}]$ with $S$ samples and sampling the angle $\Theta$ within $[-1/p,1/p]$ with $T$ samples. To mitigate the potential inaccuracies in sparse representation caused by large sampling intervals, we  ensure a minimum correlation  by setting $S \ge b_{\rm max}/\widetilde{B}_{\rm USA}$ and $T \ge 2/(pB_{\rm USA})$. Then, the representation matrix can be expressed as
\begin{align}\label{TransformMatrix}
	[\boldsymbol{W}_{\rm USA}]_{:,d} = \boldsymbol{\gamma}\left(\boldsymbol{\overline{x}},\widehat{b}_s,\widehat{\Theta}_t\right),
\end{align}
where $d = (s-1)T + t$, $\widehat{b}_s = (s-1)b_{\rm max}/S$, $\widehat{\Theta}_t = (1 +2t- T)/(pT)$, $s = 1,\cdots,S$, and $t=1,\cdots,T$.

\begin{figure}[!t]
	\centering
	\includegraphics[width=75mm]{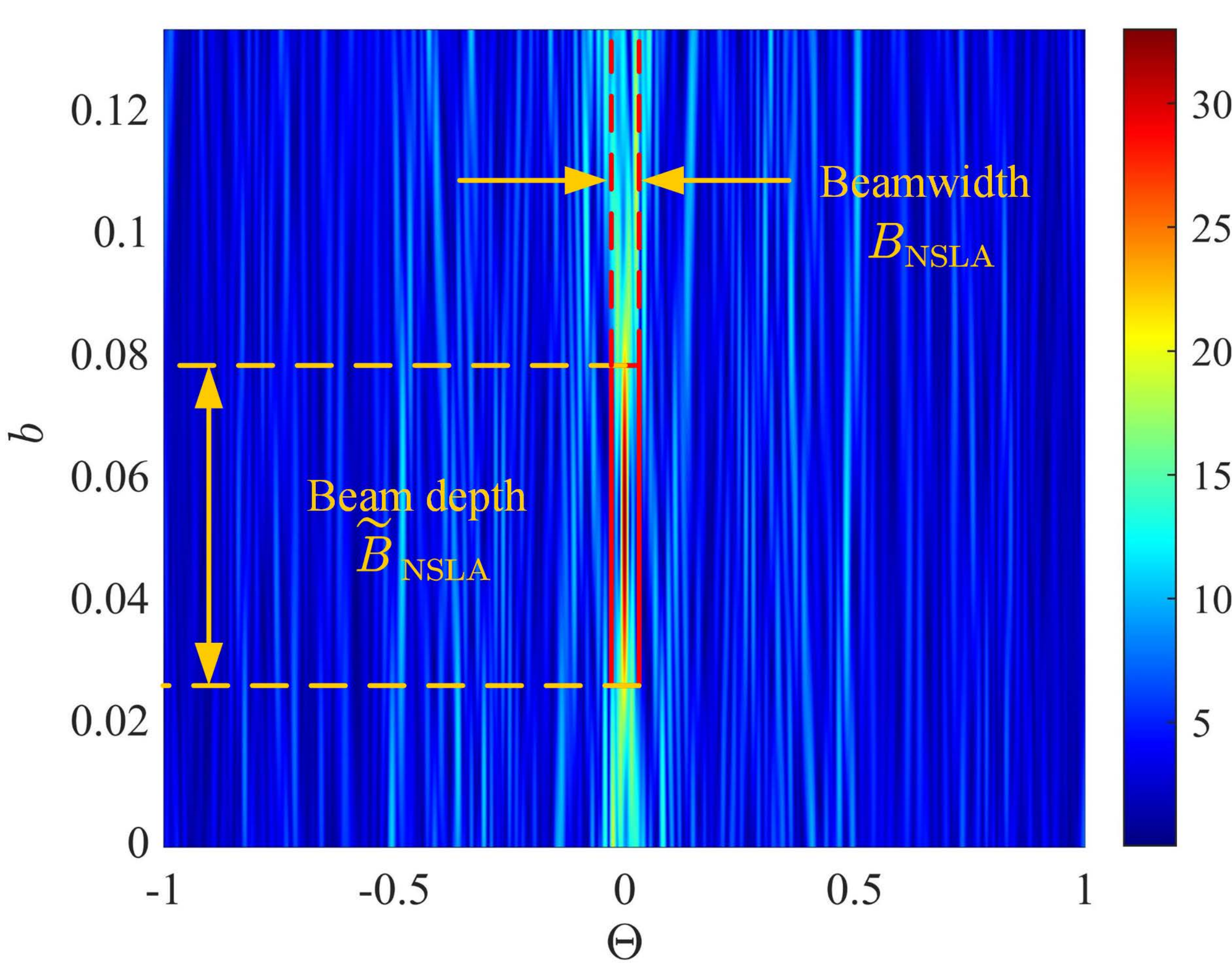}
	\caption{Illustration of $|G(\widehat{\boldsymbol{u}},b,\Theta)|$.}
	\label{NSLA}
\end{figure}

\subsubsection{Sparsity Exploration for NSAs}Based on the optimized antenna position $\widehat{\boldsymbol{x}}$, the expression of an arbitrary channel steering vector $\widehat{\boldsymbol{u}} = \boldsymbol{\gamma}(\widehat{\boldsymbol{x}},k,\Omega)$ can be obtained via \eqref{SimplifiedChannelSteeringVector2}. Then, the beam gain of $\widehat{\boldsymbol{u}}$ can be obtained via \eqref{BeamGain}. Fig.~\ref{NSLA} illustrates the absolute beam gain of $\widehat{\boldsymbol{u}}$, where we set  $N=33$, $\lambda = 0.01$~m, $p = 5$, $k = 0.05$ and $\Omega = 0$. From the figure, the beamforming of NSA can focus energy on a specific location in the SD-A domain and   does not have grating lobes. In addition, the beam gain of NSA channel steering vectors also satisfies the translation invariance, which can be easily verified  following \textbf{Proposition 1}. Therefore, the channels of NSA exhibit uniform and aperiodic sparsity in the SD-A domain. Based on this sparsity, we establish an SD-A-domain representation for the NSA channels.

First of all, we determine the coverage of mainlobes of channel steering vectors for the NSAs. Since the antenna positions of the NSAs are usually irregular, it is hard to obtain the coverage of the mainlobe via analytical methods. Alternatively, we resort to the numerical methods. Due to the translation invariance property, the shapes of the  mainlobes are the same for different channel  steering vectors. Therefore, we can determine the beamwidth and beam depth for an arbitrary channel steering vector and apply them to other channel steering vectors.  We denote the determined beamwidth and beam depth for the optimized NSA as $B_{\rm NSA}$ and $\widetilde{B}_{\rm NSA}$, respectively. Then, we quantize the channel parameters $b$ and $\Theta$ into $S$ and $T$ samples, respectively. To mitigate the potential inaccuracies in sparse representation caused by large sampling intervals, we  ensure a minimum correlation  by setting $S \ge b_{\rm max}/\widetilde{B}_{\rm NSA}$ and $T \ge 2/B_{\rm NSA}$. Then, the representation matrix can be expressed as
\begin{align}\label{NSLATransformMatrix}
	[\boldsymbol{W}_{\rm NSA}]_{:,d} = \boldsymbol{\gamma}\left(\boldsymbol{\widehat{x}},\widehat{b}_s,\widehat{\Theta}_t\right),
\end{align}
where $d = (s-1)T + t$, $\widehat{b}_s = (s-1)b_{\rm max}/S$, $\widehat{\Theta}_t = (1 +2t- T)/(pT)$, $s = 1,\cdots,S$, and $t=1,\cdots,T$.
\subsection{SD-A-Domain Orthogonal Matching Pursuit Algorithm}
Based on the sparsity of SA channels in the SD-A domain, we then propose an SDA-OMP algorithm to estimate the multiuser channels. Since the received pilots of the $K$ users are independent, we take the $k$th user as an example.

We define a residual vector $\boldsymbol{r}$ to represent the deviation between the received signal and sparse representation vector and initialize $\boldsymbol{r}$ as $\boldsymbol{r}\leftarrow \boldsymbol{y}_k$.  Denote the sparse representation dictionary as $\widetilde{\boldsymbol{W}}$, which corresponds to either $\boldsymbol{W}_{\rm USA}$ or $\boldsymbol{W}_{\rm NSA}$ based on the array configuration.  We also define an empty set $\mathcal{R}$ to keep the indices of selected vectors in the dictionary.

In the $l$th iteration, we first find the index of the vector in the dictionary, along which residual vector $\boldsymbol{r}$ has the maximum projection via
\begin{align}\label{maximumprojection}
	d^*  = \arg\max_{d \in\{1,\cdots,ST\}\backslash \mathcal{R}} \left|[\widetilde{\boldsymbol{W}}]_{:,d}^{\rm H}\boldsymbol{r}\right|^2.
\end{align}
We incorporate the selected index into $\mathcal{R}$ as 
\begin{align}\label{selectedset}
	\mathcal{R} = \mathcal{R} \cup d^*.
\end{align}
Then the corresponding selected vectors can be expressed as 
\begin{align}\label{estimatedmatrix}
	\boldsymbol{A} = \big[\widetilde{\boldsymbol{W}}\big]_{:,\mathcal{R}}.
\end{align}
With the selected vectors, we can update $\boldsymbol{r}$ by removing the projection of $\boldsymbol{y}_k$ on $\boldsymbol{A}$, which can be expressed as 
\begin{align}\label{residual}
	\boldsymbol{r} = \boldsymbol{y}_k - \boldsymbol{A}(\boldsymbol{A}^{\rm H}\boldsymbol{A})^{-1}\boldsymbol{A}^{\rm H}\boldsymbol{y}_k.
\end{align}

We iteratively operate \eqref{maximumprojection}-\eqref{residual} until the maximum number of iterations $\widetilde{L}_{k}$ is reached. Then the estimated channel can be expressed as 
\begin{align}\label{estimatedchannel}
	\widehat{\boldsymbol{h}}_k = \boldsymbol{A}(\boldsymbol{A}^{\rm H}\boldsymbol{A})^{-1}\boldsymbol{A}^{\rm H}\boldsymbol{y}_k.
\end{align}

Finally, we summarize the proposed SDA-OMP algorithm in \textbf{Algorithm~\ref{SDA-OMP}}.

\begin{algorithm}[!t]
	\caption{SD-A-Domain Orthogonal Matching Pursuit (SDA-OMP) Algorithm}
	\label{SDA-OMP}
	\begin{algorithmic}[1]
		\STATE \textbf{Input:} $\boldsymbol{y}_k$, $\widetilde{\boldsymbol{W}}$ and $\widetilde{L}_k$.		
		\STATE \textbf{Initialization:} $\boldsymbol{r}\leftarrow \boldsymbol{y}_k$, $\mathcal{R}\leftarrow \varnothing$ and  $l\leftarrow 1$.
		\WHILE{$l\le\widetilde{L}_k$}
		\STATE Obtain $d^{*}$ via \eqref{maximumprojection}.
		\STATE Update $\mathcal{R}$, $\boldsymbol{A}$, $\boldsymbol{r}$ via \eqref{selectedset}, \eqref{estimatedmatrix}, \eqref{residual}, respectively.
		\STATE $l\leftarrow l+1$.
		\ENDWHILE  
		\STATE Obtain $\widehat{\boldsymbol{h}}_{k}$ via \eqref{estimatedchannel}.
		\STATE \textbf{Output:}  $\widehat{\boldsymbol{h}}_{k}$ and $\mathcal{R}$.
	\end{algorithmic}
\end{algorithm}

\subsection{SD-A-Domain Iterative Super-Resolution Channel Estimation Algorithm}
One drawback of the SDA-OMP is the limited resolution induced by the quantization. To deal with this problem, we then propose an off-grid SDA-ISRCE algorithm, where the $k$th user is taken as an example.

The original sparse channel estimation problem can be expressed as
\begin{align}\label{zeronorm}
	\min_{\boldsymbol{\varUpsilon}_k,\boldsymbol{b}_k,\boldsymbol{\Theta}_k} \|\boldsymbol{\varUpsilon}_k\|_0,~{\rm s.t.}~\|\boldsymbol{y}_k-\boldsymbol{A}_k\boldsymbol{\varUpsilon}_k\|_2^2 \le \varepsilon,
\end{align}
where $\|\boldsymbol{\varUpsilon}_k\|_0$ denotes the number of non-zero entries in $\boldsymbol{\varUpsilon}_k$ and means the number of estimated channel paths $\widehat{L}_k$, i.e., $\boldsymbol{\varUpsilon}_k = \big[\gamma_k^{(1)},\cdots,\gamma_k^{(\widehat{L}_k)}\big]^{\rm T}$. Denote $\boldsymbol{b}_k$ and $\boldsymbol{\Theta}_k$ as the stacks of the  $\widehat{L}_k$ channel surrogate distances and channel angles, respectively.  $\boldsymbol{A}_k$ includes the corresponding channel steering vectors and can be expressed as
\begin{align}\label{stakeofcsv}
	\boldsymbol{A}_k\! =\! \left[\!\boldsymbol{\gamma}\left(\!\boldsymbol{x},[\boldsymbol{b}_k]_1,[\boldsymbol{\Theta}_k]_1\!\right),\!\cdots,\!\boldsymbol{\gamma}\left(\!\boldsymbol{x},[\boldsymbol{b}_k]_{\widehat{L}_k},[\boldsymbol{\Theta}_k]_{\widehat{L}_k}\!\right)\!\right].
\end{align}
The optimization in \eqref{zeronorm} is an NP-hard problem and is difficult to solve. An alternative approach involves using the sparse-encouraging log-sum functions, which can  efficiently approximate the $\ell_0$-norm to obtain sparse solutions~\cite{TSP16FJ}. Then, \eqref{zeronorm} can be converted~to 
\begin{align}\label{logsum}
	\min_{\boldsymbol{\varUpsilon}_k,\boldsymbol{b}_k,\boldsymbol{\Theta}_k} \sum_{l=1}^{ \widehat{L}_k}\log\big(\big|\gamma_k^{(l)}\big|^2+\delta\big),~{\rm s.t.}\|\boldsymbol{y}_k\!-\!\boldsymbol{A}_k\boldsymbol{\varUpsilon}_k\|_2^2 \le \varepsilon,
\end{align}
where $\delta>0$ is introduced to guarantee the objective is well-conditioned. By introducing a weighted factor $\varpi $, \eqref{logsum} can be converted to an unconstrained optimization problem as
\begin{align}\label{unlogsum}
&\min_{\boldsymbol{\varUpsilon}_k,\boldsymbol{b}_k,\boldsymbol{\Theta}_k} F(\boldsymbol{\varUpsilon}_k,\boldsymbol{b}_k,\boldsymbol{\Theta}_k) \nonumber \\
&~~~~~~~~~\triangleq \sum_{l=1}^{\widehat{L}_k}\log\big(|\gamma_k^{(l)}|^2+\delta\big) + \varpi  \|\boldsymbol{y}_k-\boldsymbol{A}_k\boldsymbol{\varUpsilon}_k\|_2^2.
\end{align}
Inspired by the majorization-minimization approach, \eqref{unlogsum} can be effectively solved by iteratively approximating the log-sum function with an upper-bound surrogate function  expressed as~\cite{TVT18HC}
\begin{align}\label{surrogatefunction}
	&\min_{\boldsymbol{\varUpsilon}_k,\boldsymbol{b}_k,\boldsymbol{\Theta}_k} I^{(i)}( \boldsymbol{\varUpsilon}_k,\!\boldsymbol{b}_k,\!\boldsymbol{\Theta}_k\!) \!\triangleq\!\boldsymbol{\varUpsilon}_k^{\rm H}\boldsymbol{D}^{(i)}\boldsymbol{\varUpsilon}_k\!+\!\varpi \|\boldsymbol{y}_k\!-\!\boldsymbol{A}_k\boldsymbol{\varUpsilon}_k\|_2^2,
\end{align}
where $i$ is the number of iteration, $\boldsymbol{D}^{(i)}$ is a diagonal matrix with its $l$th diagonal entry expressed as $ 1/\big(|\widehat{\gamma}_l^{(i-1)} |^2+\delta\big)$ and $\widehat{\gamma}_l^{(i-1)}$ denotes the estimation of $ \gamma_k^{(l)}$ in the $(i-1)$th iteration.

\begin{algorithm}[!t]
	\caption{SD-A-Domain Iterative Super-Resolution Channel Estimation (SDA-ISRCE) Algorithm}
	\label{ISRCE}
	\begin{algorithmic}[1]
		\STATE \textbf{Input:} $\boldsymbol{y}_k$, $\boldsymbol{x}$, $\mathcal{R}$, $\delta$, $\varpi$, $\rho$ and $\mu$.	
		\STATE \textbf{Initialization:} Initialize $\widehat{\boldsymbol{b}}_k^{(0)}$ and $\widehat{\boldsymbol{\Theta}}_k^{(0)}$ with $\mathcal{R}$, $i\leftarrow0$,
		\STATE ~~~~~~~~~~~~~~~~~~$\widehat{\boldsymbol{\varUpsilon}}_k^{(0)}  \leftarrow \sqrt{\mu}\boldsymbol{1}$ and $\widehat{\boldsymbol{\varUpsilon}}_k^{(-1)} \leftarrow \boldsymbol{0}$.
		\WHILE{$\big\|\widehat{\boldsymbol{\varUpsilon}}_k^{(i)}-\widehat{\boldsymbol{\varUpsilon}}_k^{(i-1)}\big\|_2^2 \ge \mu$}
		\STATE $i\leftarrow i + 1 $.
		\STATE Obtain $\widehat{\boldsymbol{b}}_k^{(i)}$ and $\widehat{\boldsymbol{\Theta}}_k^{(i)}$ by solving \eqref{surrogatefunction2}.
		\STATE Obtain $\widehat{\boldsymbol{\varUpsilon}}_k^{(i)}$ via \eqref{optimalfactor}.
		\IF{$\big|\widehat{\gamma}_l^{(i)}\big|^2\le\rho$}
		\STATE Discard the corresponding channel path.
		\ENDIF
		\ENDWHILE  
		\STATE $\widetilde{i}\leftarrow i$.
		\STATE Obtain $\widehat{\boldsymbol{h}}_{k}$ via \eqref{estimatedchannel2}.
		\STATE \textbf{Output:}  $\widehat{\boldsymbol{h}}_{k}$.
	\end{algorithmic}
\end{algorithm}

Note that $I^{(i)}(\!\boldsymbol{\varUpsilon}_k,\!\boldsymbol{b}_k,\!\boldsymbol{\Theta}_k\!)$ is a convex function with respect to $\boldsymbol{\varUpsilon}_k$. Given $\boldsymbol{b}_k$ and $\boldsymbol{\Theta}_k$, the optimal solution of $\boldsymbol{\varUpsilon}_k$ in the $i$th iteration can be expressed as
\begin{align}\label{optimalfactor}
\widetilde{\boldsymbol{\varUpsilon}}_k^{(i)} =  (\boldsymbol{D}^{(i)}/\varpi + \boldsymbol{A}_k^{\rm H}\boldsymbol{A}_k)^{-1}\boldsymbol{A}_k^{\rm H}\boldsymbol{y}_k.
\end{align}
 Substituting \eqref{optimalfactor} into \eqref{surrogatefunction}, we can convert \eqref{surrogatefunction} to
\begin{align}\label{surrogatefunction2}
	&\min_{\boldsymbol{b}_k,\boldsymbol{\Theta}_k} \widetilde{I}^{(i)}(\boldsymbol{b}_k,\!\boldsymbol{\Theta}_k\!) \!\triangleq -\boldsymbol{y}_k^{\rm H}\boldsymbol{A}_k\big(\boldsymbol{D}^{(i)}/\varpi + \boldsymbol{A}_k^{\rm H}\boldsymbol{A}_k\big)^{-1}\boldsymbol{A}_k^{\rm H}\boldsymbol{y}_k.
\end{align}
The optimization  problem  in \eqref{surrogatefunction2} is unconstrained and can be solved via numerical methods, such as gradient descent. We omit the details and denote the results as $\widehat{\boldsymbol{b}}_k^{(i)}$ and $\widehat{\boldsymbol{\Theta}}_k^{(i)}$.  Substituting $\widehat{\boldsymbol{b}}_k^{(i)}$ and $\widehat{\boldsymbol{\Theta}}_k^{(i)}$ into \eqref{optimalfactor}, we can obtain $\widehat{\boldsymbol{\varUpsilon}}_k^{(i)}$, which prepares for the next iteration, i.e., $\big[\widehat{\boldsymbol{\varUpsilon}}_k^{(i)}\big]_l = \widehat{\gamma}_l^{(i)}$. 

We iteratively solve \eqref{surrogatefunction} until $\big\|\widehat{\boldsymbol{\varUpsilon}}_k^{(i)}-\widehat{\boldsymbol{\varUpsilon}}_k^{(i-1)}\big\|_2^2$ is smaller than a predefined threshold $\mu$. The number of the iteration  at this point is expressed as $\widetilde{i}$. Replacing $\boldsymbol{b}$ and $\boldsymbol{\Theta}$ in \eqref{stakeofcsv} with $\widehat{\boldsymbol{b}}_k^{(\widetilde{i})}$ and $\widehat{\boldsymbol{\Theta}}_k^{(\widetilde{i})}$, we can obtain  $\widehat{\boldsymbol{A}}_k^{(\widetilde{i})}$. Then, the estimated channel can be expressed as 
\begin{align}\label{estimatedchannel2}
 \widehat{\boldsymbol{h}}_k = \widehat{\boldsymbol{A}}_k^{(\widetilde{i})}\widehat{\boldsymbol{\varUpsilon}}_k^{(\widetilde{i})}.
\end{align}

One remaining problem is how to initialize the SDA-ISRCE algorithm. During the channel estimation, we do not know the real number of channel paths and the intervals of real channel parameters. Note that the SDA-OMP can obtain estimates near the real ones. To help the convergence of the SDA-ISRCE algorithm, we take the estimation results of the SDA-OMP as the initial values.  In addition, the SDA-OMP usually obtains more channel  paths than the real ones. Throughout the iteration, we dynamically discard the estimated channel paths with smaller power than a predefined threshold $\rho$ to improve the sparsity of the estimation results.

Finally, we summarize the SDA-ISRCE algorithm in \textbf{Algorithm~\ref{ISRCE}}.

\subsection{Multiuser Beamforming}
Based on the estimated channels, we then perform the  beamforming to mitigate the MUI. In the existing literature, various  beamforming methods, such as zero-forcing, minimum mean squared error (MMSE) and weighted MMSE,  have been developed. Taking the MMSE as an example, we design the digital beamformer as
\begin{align}
	\widehat{\boldsymbol{F}} = \widehat{\boldsymbol{H}}^{\rm H}\left(\widehat{\boldsymbol{H}}\widehat{\boldsymbol{H}}^{\rm H} + \sigma^2\boldsymbol{I}_K\right)^{-1},
\end{align} 
where $ \widehat{\boldsymbol{H}} \triangleq \left[ \widehat{\boldsymbol{h}}_1, \widehat{\boldsymbol{h}}_2,\cdots, \widehat{\boldsymbol{h}}_K\right]$. To satisfy the total transmit power constraint,  we normalize $\widehat{\boldsymbol{F}}$ as $\overline{\boldsymbol{F}} = \sqrt{K}\widehat{\boldsymbol{F}}/\big\|\widehat{\boldsymbol{F}}\big\|_{\rm F}^2$.

\begin{figure}[!t]
	\begin{center}
		\includegraphics[width=76mm]{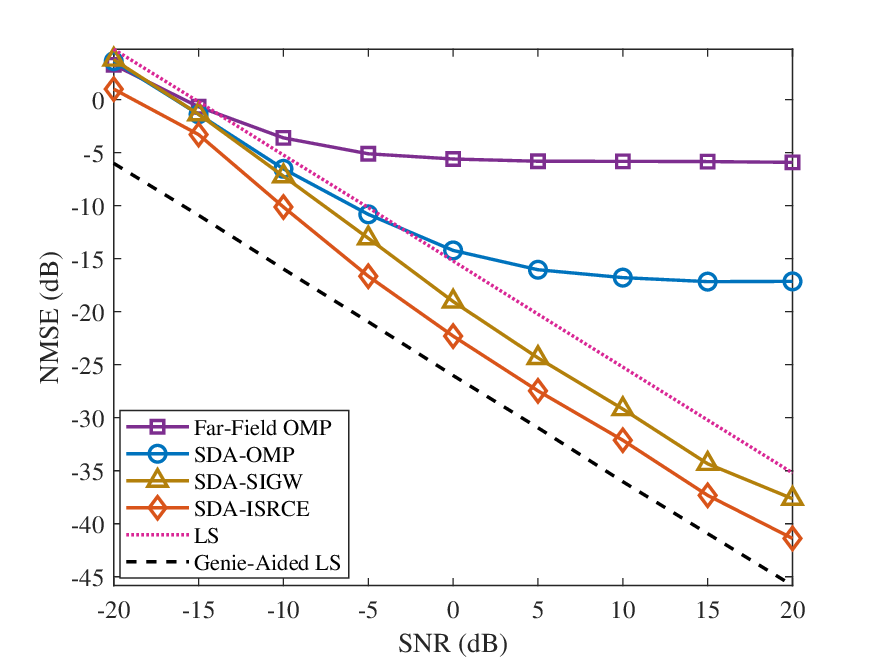}
	\end{center}
	\caption{Comparisons of different methods in terms of the NMSE for different SNRs.}
	\label{CE}
\end{figure}

\section{Simulation Results}\label{SimulationResults}
Now, we evaluate the performance of the proposed near-field multiuser communications based on the USAs and NSAs. The BS employs an SA with $33$ antennas to serve $K$ users. The communication systems operate at the carrier frequency of $30$~GHz, which corresponds to the carrier wavelength of $0.01$~m. The channel between the BS and the $k$th user contains one line-of-sight path and two non-line-of-sight paths, where the Ricean K-factor is denoted as $\kappa$. The total power of the users is normalized as $K$ and the SNR is calculated as $10\log_{10}(\sigma^{-2})$. For the SCA-APO algorithm, we set the maximum number of iterations as $Q = 100$. We also perform the multiuser communications with the uniform circular array (UCA) in \cite{TWC23WZD} and the conventional HULA, which are adopted as benchmarks.  For a fair comparison, the UCA and HULA have the same simulation configurations as the proposed USA and~NSA.

In Fig.~\ref{CE}, we compare the normalized mean squared error (NMSE) of different channel estimation methods for different SNRs. The channel angles  distribute randomly within $[-\sqrt{3}/2,\sqrt{3}/2]$, while the channel distances  distribute randomly within $[10,100]$~m. The Ricean K-factor is set as $\kappa = -10$~dB. Since the NSA has the most complicated structure among all antenna configurations, we take the channel estimation of the NSA as an example and set the array sparsity factor as $p = 10$.  We adopt the far-field OMP and least squares (LS) as the benchmarks. The SD-A-domain simultaneous iterative gridless weighted (SDA-SIGW) algorithm, which is developed from the polar-domain SIGW in \cite{Tcom22CMH} is also adopted as a benchmark. The genie-aided LS method, which assumes that the channel angles and distances are known and then estimates the channel gains with the LS method, is employed as the lower bound~\cite{Tcom21LW}. From the figure, the far-field OMP has the worst performance among all the methods. This is because the significantly expanded array aperture of NSA leads to the inaccuracy of the far-field assumption and the far-field channel estimation method is not effective in this condition. In addition, the SDA-ISRCE outperforms the far-field OMP and SDA-OMP thanks to its off-grid characteristics. The SDA-ISRCE outperforms the LS due to the exploitation of the SD-A-domain sparsity. The SDA-ISRCE outperforms the SDA-SIGW because the former considers both the channel sparsity and data fitting accuracy while the latter only considers the data fitting accuracy.  Moreover, the SDA-ISRCE has a small gap with the genie-aided  LS method, which verifies the effectiveness of the SDA-ISRCE.

\begin{figure}[!t]
	\begin{center}
		\includegraphics[width=76mm]{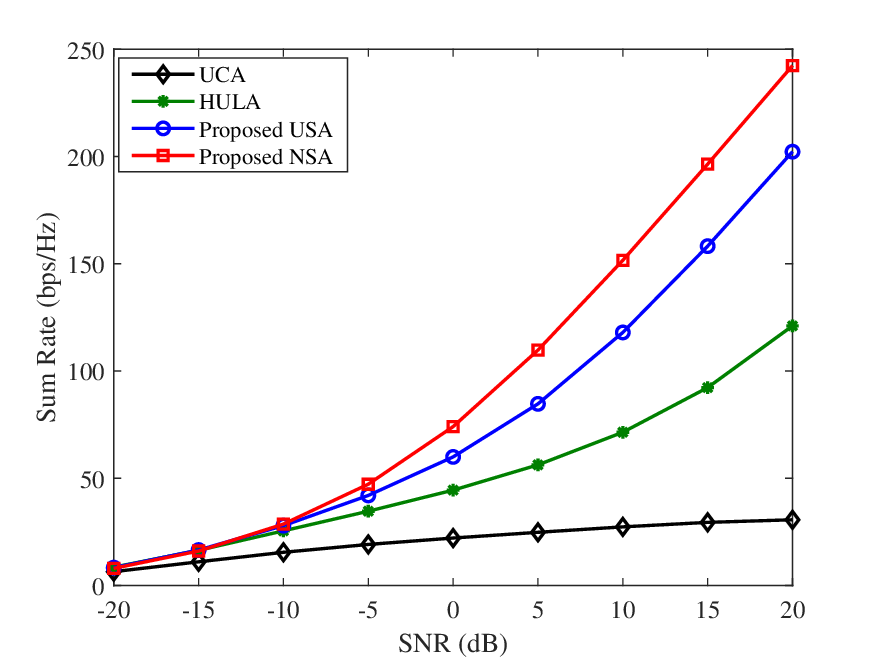}
	\end{center}
	\caption{Comparisons of different arrays in terms of sum rate for different SNRs.}
	\label{DSNR}
\end{figure}

In Fig.~\ref{DSNR}, we evaluate the performance of multiuser communications for different kinds of arrays, considering different SNRs. The sum rate is employed as the metric for evaluating the performance  of multiuser communications, following the works in \cite{SPM14BE}. Specifically, we set $\kappa = -20$~dB and  $K=28$. The array sparsity factor, the distribution of the channel angles and the distribution of the channel distances are the same as those in Fig.~\ref{CE}. From the figure, when the SNR is low, e.g., less than $-10$~dB, the four arrays achieve similar performance. In fact, this similarity arises from the substantial  deterioration caused by the noise, which impacts the effectiveness of each method. However, at high SNRs, the four arrays have different performance. Specifically, the sum rate of the UCA is notably  lower than those of the other three arrays. This is because the adoption of a circular array configuration for a fixed number of antennas results in a significantly reduced array aperture compared to the other three arrangements.  This diminished array aperture, in turn, adversely affects angle resolution and exacerbates the MUI. In contrast, the proposed USA  and  NSA  demonstrate significantly higher sum rate than the UCA and HULA. This improvement can be attributed to the larger array apertures of the former two arrays. The expanded array apertures of the USA and NSA enable the exploitation of near-field effects, which increases the spatial resolution and consequently increases the sum rate in multiuser communications. Notably, despite similar array apertures, the NSA outperforms the USA. This superior performance is due to NSA's ability of removing the grating lobes, thereby reducing MUI and increasing the sum rate.


\begin{figure}[!t]
	\begin{center}
		\includegraphics[width=76mm]{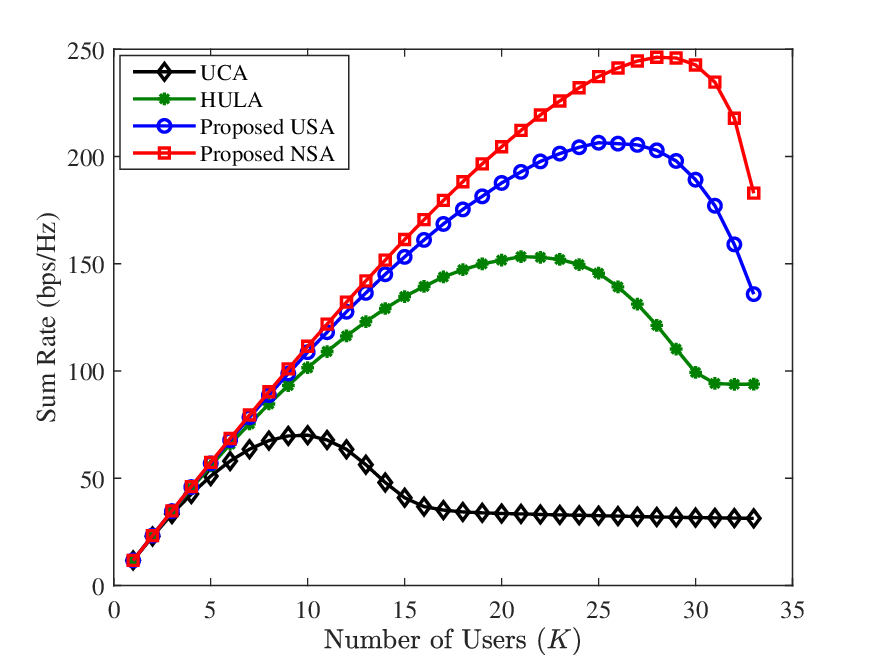}
	\end{center}
	\caption{Comparisons of different arrays in terms of sum rate for different numbers of users.}
	\label{DK}
\end{figure}

\begin{figure}[!t]
	\begin{center}
		\includegraphics[width=76mm]{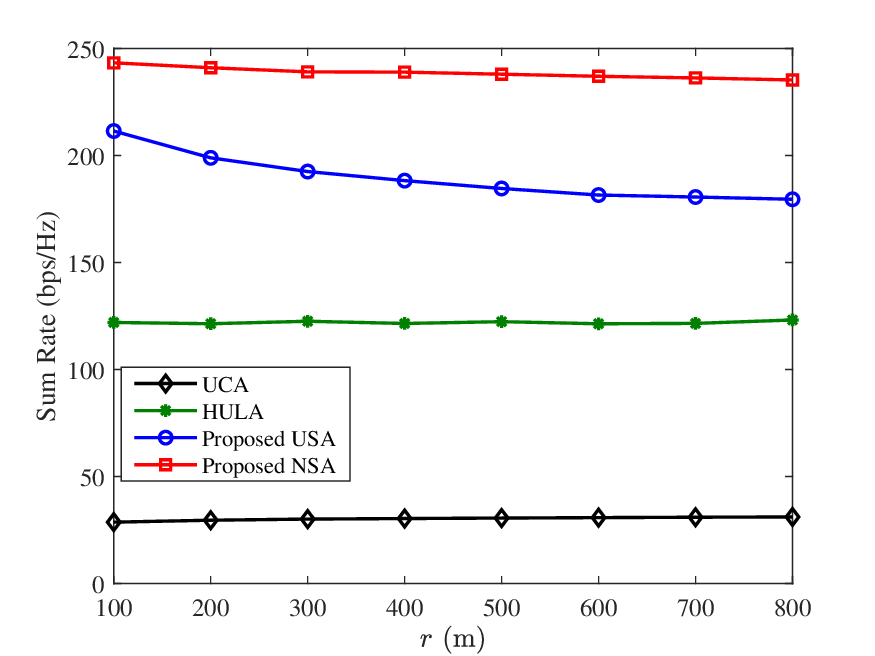}
	\end{center}
	\caption{Comparisons of different arrays in terms of sum rate for different distances.}
	\label{DD}
\end{figure}

In Fig.~\ref{DK}, we compare the sum rates of multiuser communications for different kinds of arrays, considering different numbers of users.  We fix  the SNR as $20$~dB. The array sparsity factor, the distribution of the channel angles and the distribution of the channel distances are the same as those in Fig.~\ref{DSNR}. When the number of users is small, e.g., $K\le5$, all the four arrays can provide enough spatial DoF to separate multiple users. As a result,  the four arrays have similar performance in terms of sum rate. With the increase of $K$, the sum rates of the four arrays all initially increase  and then decrease. This trend is attributed to the detrimental impact of strong  MUI  on the performance of multiuser communications when dealing with a larger number of users. Notably, the four arrays achieve  the maximum sum rates at different user counts. Specifically,  the UCA, HULA, USA and NSA achieve their peak sum rates when the number of users  equals $10$, $21$, $25$ and $28$, respectively. This observation  indicates that the sparse arrays have the potential to serve more users than the conventional HULA.

\begin{figure}[!t]
	\begin{center}
		\includegraphics[width=76mm]{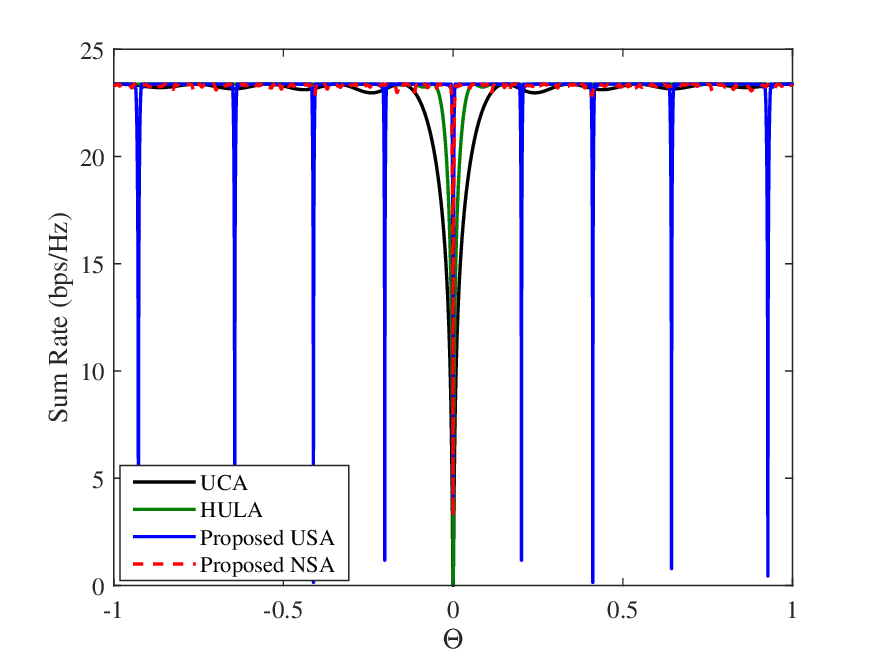}
	\end{center}
	\caption{Comparisons of different arrays in terms of sum rate for of varying angles.}
	\label{VaryAngles}
\end{figure}

In Fig.~\ref{DD}, we compare the sum rates of multiuser communications for different kinds of arrays, considering different distances. The channel distances distribute randomly within $[10,r]$, where $r$ varies from $100$~m to $800$~m. The number of users is fixed as $K = 28$. The array sparsity factor, the distribution of the channel angles and the SNR are the same as those in Fig.~\ref{DK}. From the figure, the proposed USA and NSA perform better than the UCA and HULA due to their larger array apertures. The UCA and HULA show similar performance for different distances, as their channels are predominantly influenced by the far-field components, leading to the nearly invariant performance with changing distances. Conversely, the performance of the proposed USA and NSA diminishes as  distance increases. This decline is attributed to the channels shifting towards far-field characteristics as the distance increases. The consequent reduction in spatial resolution deteriorates the performance of multiuser communications for these sparse arrays. Note that the Rayleigh distance of the considered system is 512 m. The proposed arrays demonstrate superior  performance to the existing ones in different distances, ranging from 100 m to 800 m. Therefore, the proposed arrays outperform the existing ones in both near and far fields.

In Fig.~\ref{VaryAngles}, we compare the sum rates of two users for different kinds of arrays, considering variations in channel angles. First, we fix the x-axis and y-axis coordinates of the first user as $0$~m and $100$~m, respectively, corresponding to the channel angle $\Theta =0$ and channel distance of $100$~m. Subsequently, we maintain the channel surrogate distance of the second user the same as the first user while varying its channel angle. For simplicity, we assume that the channels for both users only contain the line-of-sight path. From the figure, the USA has ten nulls, which are caused by the grating lobes.  Furthermore, the widths of the nulls for both USA and  NSA  are considerably smaller than those of the  UCA and  HULA. This discrepancy indicates that USA and NSA have higher angle resolution due to their larger array apertures.



\begin{figure}[!t]
	\begin{center}
		\includegraphics[width=76mm]{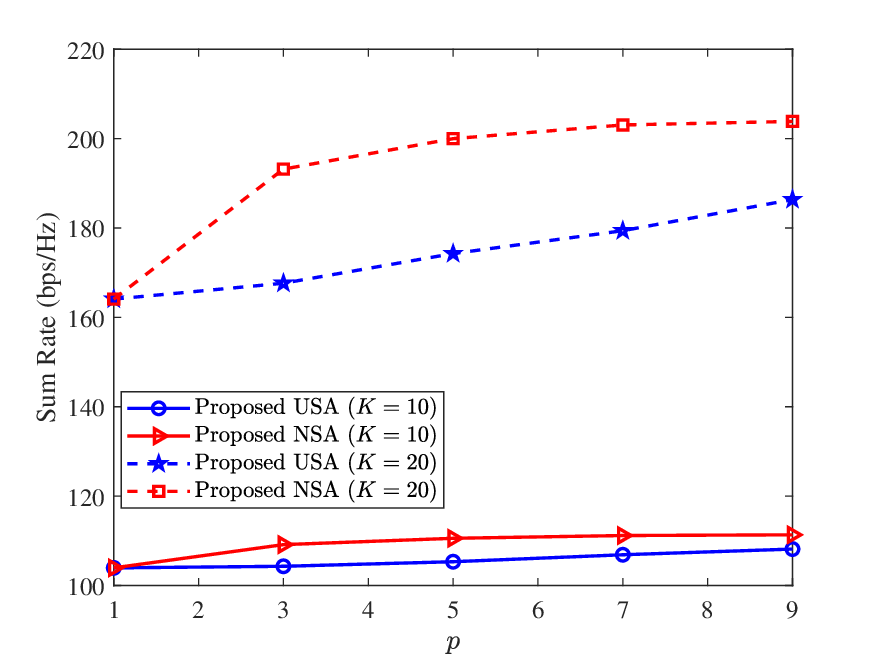}
	\end{center}
	\caption{Comparisons of different arrays in terms of sum rate for different antenna spacings.}
	\label{ChangeWithSpace}
\end{figure}

In Fig.~\ref{ChangeWithSpace},  we compare the sum rates of multiuser communications for different antenna spacings.  From the figure, the NSA outperforms the USA due to the mitigation of grating lobes. Furthermore, the sum rates of both USA and NSA exhibit an increasing trend with $p$. When $K=10$, the sum rates show a gradual increase, whereas a significant increase is observed when $K=20$. This is because antenna arrays with smaller apertures can  provide sufficient spatial resolution to separate multiple users when the number of users is small. Conversely, as the number of users increases, arrays with smaller apertures cannot provide adequate spatial resolution. This observation underscores the effectiveness of sparse arrays in enhancing multiuser communication performance by augmenting the array aperture.

\section{Conclusion}\label{Conclusion}
In this paper, near-field multiuser communications based on SAs have been considered. First, for the USAs, the beam gains of channel steering vectors have been analyzed. The NSAs have been investigated to mitigate the high MUI from the grating lobes of USAs. To maximize the sum rate of near-field multiuser communications,  the antenna positions of the NSAs have been optimized and a successive convex approximation-based antenna position optimization algorithm has been proposed. Moreover, we have found that channels of both USAs and NSAs show uniform sparsity in the SD-A  domain. Then, an on-grid SDA-OMP algorithm and an off-grid  SDA-ISRCE algorithm have been proposed. Simulation results have demonstrated the superior performance of the proposed methods.

For future works, we will investigate the beam training for near-field communications based on SAs. In addition, we will exploit the spatial correlation in Rician fading channel model to improve the performance of near-field communications, following the works in \cite{Tcom2019EB} and \cite{TSP21PD}. Furthermore, we will explore the potential of the SAs in enhancing the capacity of near-field communications.	

\bibliographystyle{IEEEtran}
\bibliography{IEEEabrv,IEEEexample}

\end{document}